\documentclass[twocolumn]{aastex631}

\usepackage{comment}
\usepackage{amssymb}
\usepackage{amsmath}
\usepackage{xspace}
\usepackage{enumitem}
\usepackage{graphicx}
\usepackage{rotating}
\usepackage{color}
\usepackage{float}
\usepackage{gensymb}
\usepackage{multirow}
\usepackage{threeparttable}
\received{2024 April 23}
\accepted{2024 July 30}
\submitjournal{PASP}

\newcommand {\Tef}   {$T_{\mathrm {eff}}$}
\newcommand {\K}     {{$\mathrm{~K}$}}
\newcommand {\lgg}   {$\mathrm \log ~g$}

\newcommand {\feh}   {[Fe/H]}

\begin{document}

\title{SPECtrophotometer for TRansmission spectroscopy of exoplanets (SPECTR)}

\author[0009-0005-5145-5165]{Yeon-Ho Choi}
\affiliation{Department of Astronomy and Space Science, University of Science and Technology, Korea, 217 Gajeong-ro, Yuseong-gu, Daejeon 34113, Republic of Korea}
\affiliation{Korea Astronomy and Space Science Institute,
776 Daedeok-daero, Yuseong-gu, Daejeon 34055, Republic of Korea}

\author[0000-0003-1544-8556]{Myeong-Gu Park}
\affiliation{Department of Astronomy and Atmospheric Sciences, Kyungpook National University, 80 Daehak-ro, Buk-gu, Daegu 41566, Republic of Korea}

\author{Kang-Min Kim}
\affiliation{Korea Astronomy and Space Science Institute,
776 Daedeok-daero, Yuseong-gu, Daejeon 34055, Republic of Korea}

\author[0000-0001-8969-0009]{Jae-Rim Koo}
\affiliation{Earth Environment Research Center, Kongju National University, 56 Gongjudaehak-ro, Gongju-si, Chungcheongnam-do 32588, Republic of Korea}

\author[0009-0008-2592-9437]{Tae-Yang Bang}
\affiliation{Department of Astronomy and Atmospheric Sciences, Kyungpook National University, 80 Daehak-ro, Buk-gu, Daegu 41566, Republic of Korea}

\author{Chan Park}
\affiliation{Korea Astronomy and Space Science Institute,
776 Daedeok-daero, Yuseong-gu, Daejeon 34055, Republic of Korea}

\author{Jeong-Gyun Jang}
\affiliation{Korea Astronomy and Space Science Institute,
776 Daedeok-daero, Yuseong-gu, Daejeon 34055, Republic of Korea}

\author{Inwoo Han}
\affiliation{Korea Astronomy and Space Science Institute,
776 Daedeok-daero, Yuseong-gu, Daejeon 34055, Republic of Korea}

\author{Bi-Ho Jang}
\affiliation{Korea Astronomy and Space Science Institute,
776 Daedeok-daero, Yuseong-gu, Daejeon 34055, Republic of Korea}

\author{Jong Ung Lee}
\affiliation{Division of Energy and Optical Technology Convergence, Cheongju University, 298 Daeseong-ro, Cheongwon-gu, Cheongju 28371, Republic of Korea}

\author{Ueejeong Jeong}
\affiliation{Korea Astronomy and Space Science Institute,
776 Daedeok-daero, Yuseong-gu, Daejeon 34055, Republic of Korea}

\author{Byeong-Cheol Lee}
\affiliation{Department of Astronomy and Space Science, University of Science and Technology, 217 Gajeong-ro, Yuseong-gu, Daejeon 34113, Republic of Korea}
\affiliation{Korea Astronomy and Space Science Institute,
776 Daedeok-daero, Yuseong-gu, Daejeon 34055, Republic of Korea}

\correspondingauthor{Byeong-Cheol Lee}
\email{bclee@kasi.re.kr}

\begin{abstract}

The SPECtrophotometer for TRansmission spectroscopy of exoplanets (SPECTR) is a new low-resolution optical (3800~\AA~- 6850~\AA) spectrophotometer installed at the Bohyunsan Optical Astronomy Observatory (BOAO) 1.8 m telescope. SPECTR is designed for observing the transmission spectra of transiting exoplanets. Unique features of SPECTR are its long slit length of 10 arcminutes which facilitates observing the target and the comparison star simultaneously, and its wide slit width to minimize slit losses. SPECTR will be used to survey exoplanets, such as those identified by the Transiting Exoplanet Survey Satellite (\textit{TESS}), providing information about their radii across the wavelength range. In this paper, we present the design of SPECTR and the observational results of the partial transit of HD~189733~b and a full transit of Qatar\mbox{-}8~b. Analyses show the SPECTR's capability on the white light curves with an accuracy of one ppt. The transmission spectrum of HD~189733~b shows general agreement with previous studies.


\end{abstract}




\section{Introduction} \label{sec:intro}

The \textit{Kepler/K2} and the \textit{TESS} space missions so far have confirmed about 3,800 exoplanet candidates, 67\% of which are confirmed exoplanets\footnote[1]{https://exoplanetarchive.ipac.caltech.edu/}. 
There are about 10,000 candidates yet to be confirmed. While these successful missions have discovered a substantial number of exoplanets, further follow-up observations, such as precise radial velocity measurements, are still necessary to determine their masses. Furthermore, there has been a growing importance placed on studying the atmospheric characteristics of exoplanets, such as atmospheric composition and the presence of clouds using transmission and thermal emission spectroscopy methods.

In particular, transmission spectroscopy, employed for transiting exoplanets, stands out as an exceptional tool for investigating the upper atmospheres of these planets. This method has been widely used with both space- and ground-based telescopes over a range of spectral resolutions. Space telescopes such as the \textit{Hubble} Space Telescope \citep[HST, e.g.,][]{cha02, vid03, dem13, sin16, arc18} and the \textit{JWST} \citep[e.g.,][]{JWST2023}, along with ground-based telescopes such as the Very Large Telescope \citep[VLT, e.g.,][]{Nikolov_FORS2, Borsa_ESPRESSO}, the Gemini Telescope \citep[e.g.,][]{Huitson_GMOS, Line_IGRINS}, and the Keck Telescope \citep[e.g.,][]{Crossfield_MOSFIRE, Kirk_NIRSPEC}, and programs such as the ACCESS program with the Magellan Telescope \citep[e.g.,][]{Rackham17, Espinoza19, Bixel19, McGruder2020, Allen2022} and the LRG-BEASTS program with the William Herschel Telescope \citep[WHT, e.g.,][]{Kirk18, Alderson20, Ahrer22} have successfully studied the atmospheric characteristics of exoplanets. Until recently, such studies were limited to space- and ground-based large telescopes due to the faintness of the stars.

However, \textit{TESS} has been particularly effective in discovering exoplanets orbiting bright stars. The atmospheres of these exoplanets can be studied using smaller telescopes like the BOAO 1.8 m telescope. This is because the 1.8 m telescope at BOAO, with its limited collecting area, is best suited for observing bright stars, which are also the primary targets of \textit{TESS} due to its own small telescope size.

With these opportunities, we aimed to develop a spectrophotometer capable of studying the atmospheric characteristics of the \textit{TESS} targets with the Bohyunsan Optical Astronomy Observatory (BOAO) 1.8 m telescope located in South Korea.

Korea Astronomy and Space Science Institute (KASI) and Kyungpook National University (KNU) initiated the development of the SPECtrophotometer for TRansmission spectroscopy of exoplanets (SPECTR). 
Considering the given situation and budget, we concluded that a long-slit spectrophotometer type is most suitable for studying the atmospheric characteristics of exoplanets at BOAO. While a long-slit spectrograph was operating at BOAO, it was not suitable for performing transmission spectroscopy due to its small field of view (FOV, 3.3\arcmin $\times$ 3.3\arcmin) and the absence of an imaging mode. The FOV determines the likelihood of finding a proper comparison star, necessary to perform precision differential spectrophotometry from the ground, due to the variability of Earth's atmosphere. Also, an imaging mode is required to confirm the position angle between the target and the comparison star to ensure that both stars are located within the same slit. 

In this paper, we describe the optical design of SPECTR and its first observational results.

\section{SPECTR Design and System}
    \subsection{Optical design}
    SPECTR is designed to fit into the BOao Echelle Spectrograph (BOES) Cassegrain Interface Module \citep[CIM,][]{Kim2002}. 
    The CIM, attached to the Cassegrain focus of the BOAO 1.8 m telescope, includes a slit assembly, a slit-monitoring system, and a calibration lamp system. 
    The CIM is designed to facilitate high-resolution observation with BOES and medium-resolution observations with the Long Slit Spectrograph (LSS). BOES is a fiber-fed echelle spectrograph capable of providing precise radial velocity measurements \citep{Han2010}. BOES covers a spectral range from 3,500~\AA~to 10,100~\AA. Its spectral resolution varies from 30,000 to 90,000, depending on the fiber diameter, which ranges from 80 microns to 300 microns.
    
\begin{figure} \centering
    \includegraphics[width=0.47\textwidth]{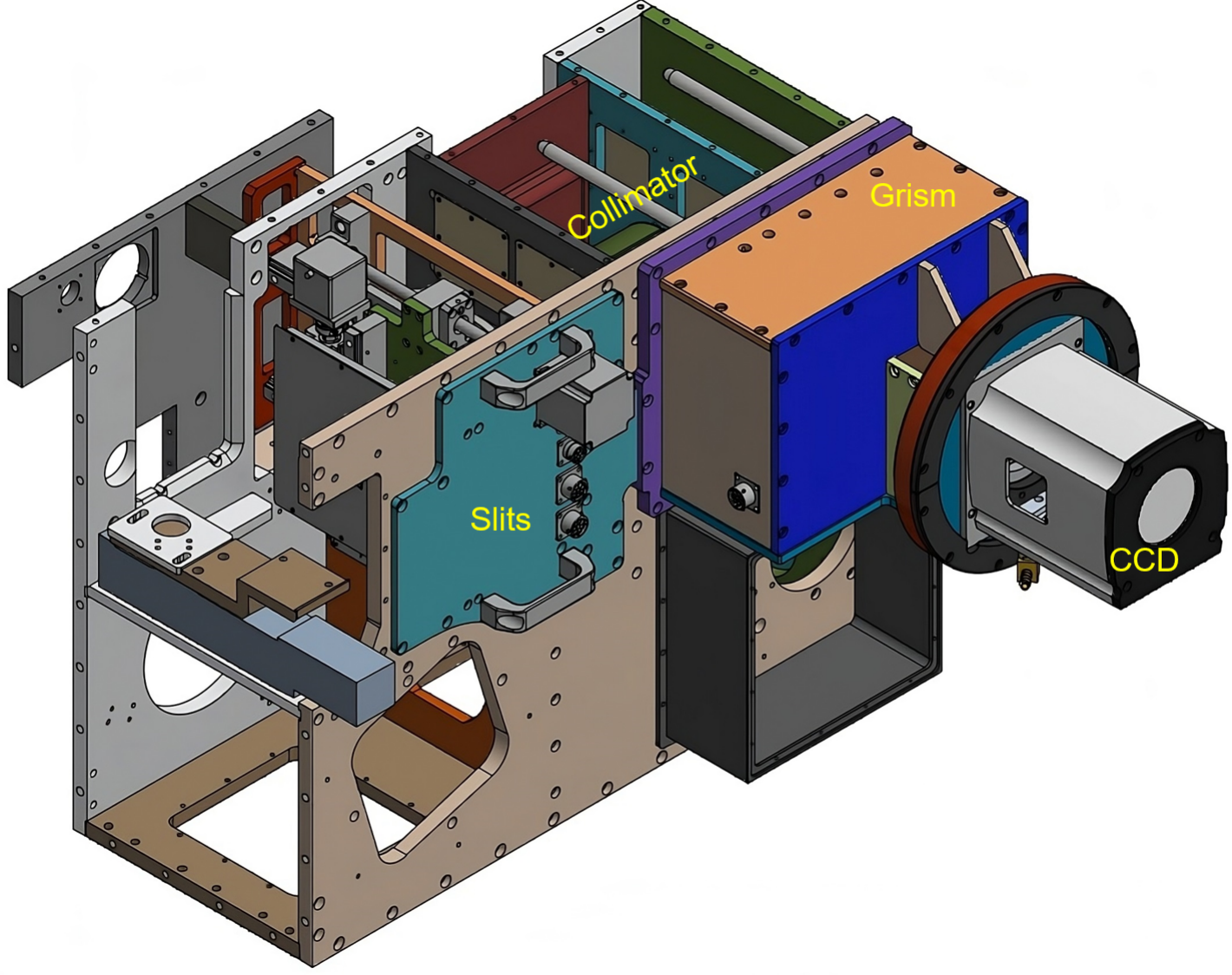}
    \caption{Schematics of the CIM inside of which is divided into several compartments by walls. SPECTR fits into the CIM as two removable cartridges: one containing the slit assembly and the other containing the collimator assembly, grism assembly, and imaging assembly.}
    \label{fig:CIM}
    \end{figure}

\begin{figure} \centering
    \includegraphics[width=0.44\textwidth]{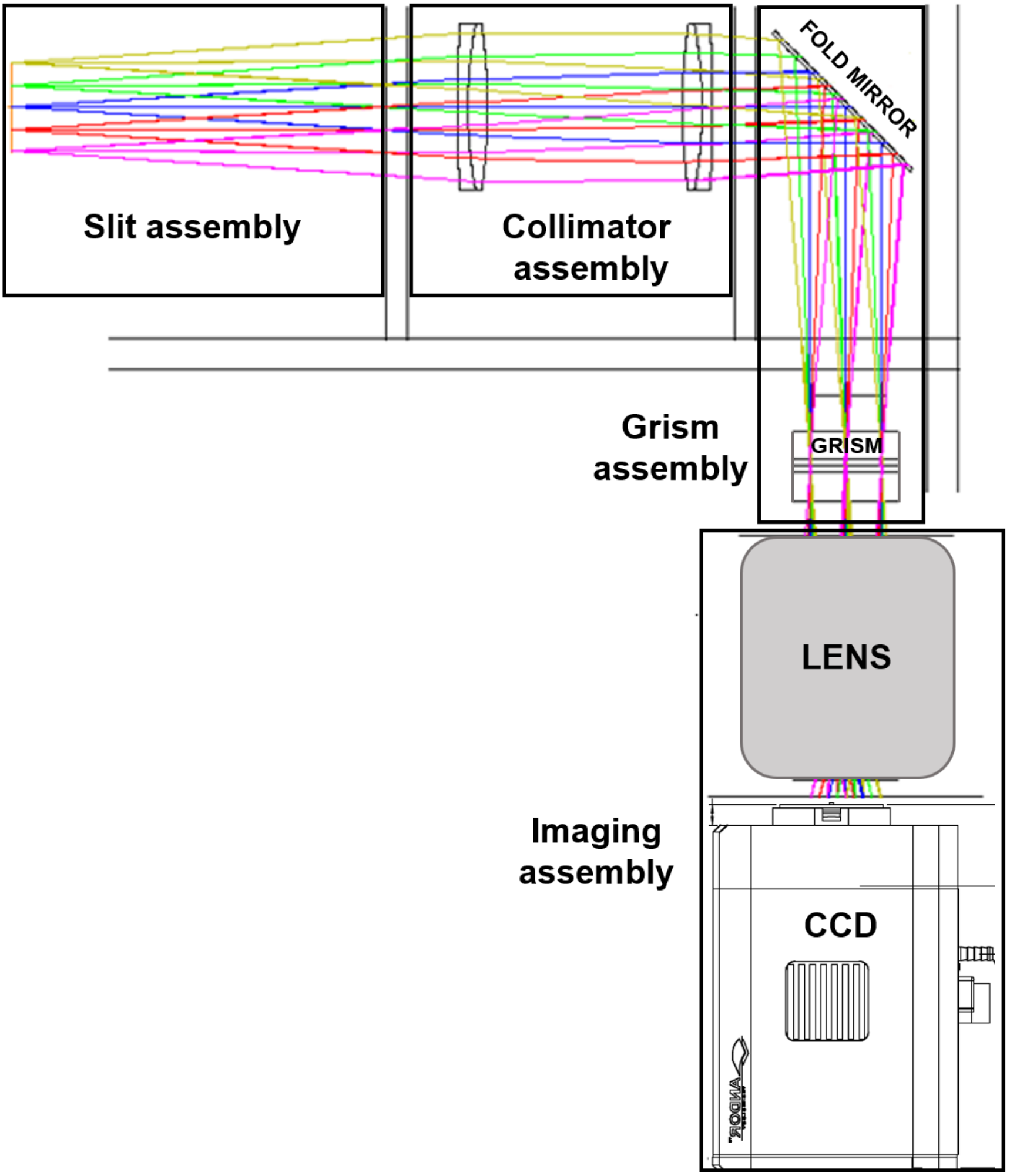}
    \caption{Optical layout of SPECTR containing four major parts: slit assembly, collimator assembly, grism assembly, and imaging assembly.}
    \label{fig:SPECTR_optics}
    \end{figure}

    The original design of SPECTR had a straight optical path to enable the imaging mode utilizing the volume phase holographic grating (VPHG). 
    However, due to the limited space and compartments formed by walls inside the CIM as shown in Figure~\ref{fig:CIM}, it was changed to an L-shaped design using a flat mirror. Figure~\ref{fig:SPECTR_optics} shows the optical layout of SPECTR. SPECTR consists of a slit assembly, a collimator assembly, a grism assembly, and an imaging assembly.

    The optical design of an L-shape can cause an imbalance at the center of mass in the CIM. To address this, we conducted simulations to test the amount of deflection due to changing gravity. We confirmed that the maximum deflection of the outer part of SPECTR at 1 g was 22 microns, which corresponded to a 2.6 \AA~shift in the spectral axis. Given that we use 400 \AA~bins for the spectroscopic light curves, this 2.6 \AA~shift is minor and will not severely affect the results.
    
    \subsection{Slit assembly}
    SPECTR has a wavelength coverage from 3,800~\AA~to 6,850~\AA ~and a spectral resolution (R) ranging from 200 to 1,300 depending on the slit width. Details of the spectral resolutions for each slit are listed in Table~\ref{slit}.
    SPECTR contains eight slits of different widths, in addition to one pinhole for testing and one blank space for imaging to check the positions of the target and the comparison stars. Slit \#1 is mainly used for taking the spectrum from the FeAr lamp for wavelength calibration and slits from \#2 to \#4 for general long-slit spectroscopy. Since transit observations require minimizing the light loss which may arise due to a finite slit width, we use slits from \#5 to \#8, much wider than the typical seeing at BOAO. 
    
    The slit plates are made of stainless steel (STS), measuring $60$ mm~$\times$ $22$ mm, and they are coated with SiO${_2}$ after metal polishing. The slits are manufactured by wire electric discharge machining and have the same physical length of 42 mm, which corresponds to 10 arcminutes in the FOV. 
    The blank space has the physical dimensions of $46$ mm~$\times$ $20$ mm.
    
    \begin{table}
    \centering
    \caption{Components of slit assembly}
    \label{slit}
    \begin{tabular}{cccccc}
    \hline\hline
    Slit & Slit dimension & Slit width & R & Note  \\
    Number & [$\mu$m]& [\arcsec]&[$\lambda / \Delta\lambda$] & \\
    \hline
    \#0 & 70 & 1.0 &  & Pinhole  \\
    \#1 & 100 & 1.4 & 1000 - 1450 & \\
    \#2 & 140 & 2.0 & 790 - 1150 & \\
    \#3 & 200 & 2.8 & 600 - 860 & \\
    \#4 & 280 & 4.0 & 430 - 630 & \\
    \#5 & 420 & 6.0 & 325 - 470 & \\
    \#6 & 630 & 9.0 & 225 - 320 & \\
    \#7 & 1050 & 15.0 & 135 - 190 & \\
    \#8 & 1750 & 25.0 & 80 - 110 & \\
    \#9 & \multicolumn{2}{c}{46 mm $\times$ 22 mm}  & --- & Imaging \\
    \hline
    \end{tabular}
    \end{table}

\begin{figure}[t] \centering
    \includegraphics[width=0.95\linewidth]{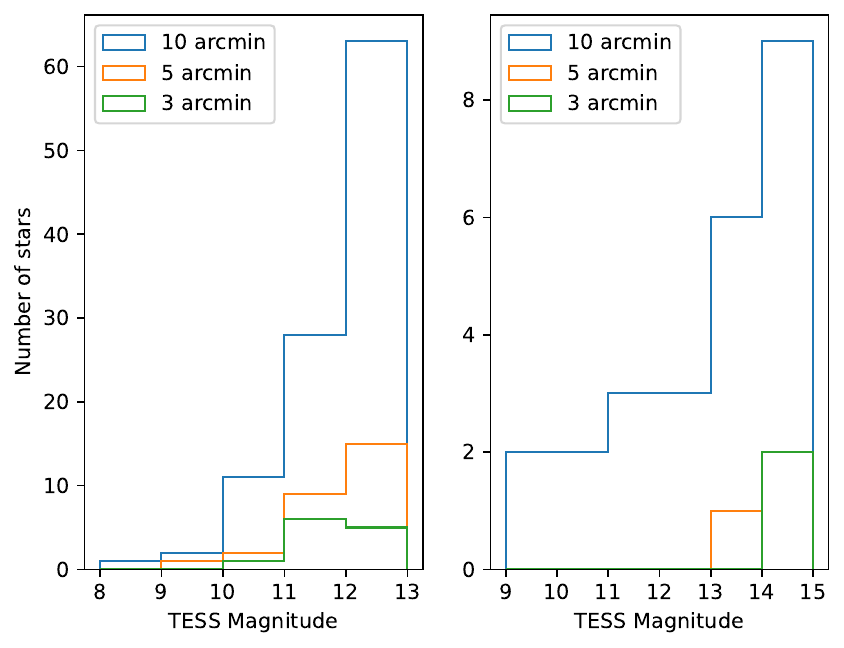}
    \caption{Number of stars brighter than TESS magnitude 13 and 15 in 1 magnitude bins for slit FOVs of 3, 5, and 10 arcminutes. The left panel shows the number of stars at a galactic latitude of 0\degree, while the right panel shows the number of stars at a galactic latitude of 90\degree. For the galactic latitude of 0\degree, galactic longitude was chosen to represent an average number of stars between 0\degree and 180\degree. The corresponding RA and DEC are [21:12:01.05 +48:19:46.71]. Similarly, for the galactic latitude of 90\degree, the corresponding RA and DEC are [12:51:26.28 +27:07:41.70]. Both positions were chosen for their accessibility from BOAO. 
    }
    \label{fig:stardensity}
    \end{figure}

    In ground-based transmission spectroscopy, it is essential to observe a comparison star simultaneously with the target to correct for the effects of the changing of the Earth's atmosphere. A longer slit length will increase the probability of finding a suitable comparison star for the target. The LSS is already installed for low-resolution spectroscopy, but its FOV covers only one-third of that of SPECTR, limiting its capability to capture comparison stars simultaneously.
    In addition, the slit of SPECTR can be rotated to match the position angle of the two stars.
    To arrive at the optimal length of the slit within the optical and physical limitations of the telescope and the CIM, we estimated the number of suitable comparison stars available within a given FOV using the \textit{TESS} Input Catalog \citep[TIC,][]{Stassun2018}. During test observations, we found that a proper comparison star should have a difference in photometric magnitude of at most 1 magnitude in the V band and have colors or spectral types similar to the target. The TIC provides \textit{TESS} magnitudes for about 471 million bright stars. Moreover, \textit{TESS} provides light curves of stars, that can be used to check the photometric variability of potential comparison stars.
    Figure~\ref{fig:stardensity} shows the number of stars with magnitudes within the interval of 1 magnitude for slit FOVs of 3, 5, and 10 arcminutes.

    \subsection{Collimator assembly}
    A collimator with an effective focal length of 280 mm was designed to change the f/8 beam from the 1.8 m telescope to a 35 mm diameter at the SPECTR grism. The collimator consists of two symmetric achromatic doublet lenses, which collimate the incident light towards the grism. The doublet lenses are fixed in place using two spacers, axial springs, radial springs, and retainers.
    \bigskip
    \subsection{Grism assembly}
    The collimated beam is reflected by the folded mirror toward the grism and dispersed toward the imaging lens. The folded mirror has a physical dimension of $94$ mm~$\times$ $66$ mm and is coated with enhanced aluminum for durability and reflectance with SPECTR's wavelength coverage.

    The grism assembly houses the grism. The grism consists of a VPHG and a wedge prism. The detailed parameters of the VPHG are given in Table~\ref{vphg}. 
    It is manufactured by Wasatch Photonics, USA\footnote{https://wasatchphotonics.com/}. Two N-BK7 wedge prisms with an apex angle of 25.3\degree~are glued to either side of the VPHG which together produces a straight optical path at 5261~\AA. The grism is installed on the X-stage with a step motor and therefore can be moved the grsim away from the light path for the imaging mode.

    \begin{table}
    \centering
    \caption{Specifications of the SPECTR VPHG}
    \label{vphg}
    \begin{tabular}{lc}
    \hline\hline
    Parameter & Value  \\
    \hline
    Grating blaze angle & 20\degree \\
    Grating groove frequency & 900 lines/mm \\
    Grating blaze wavelength ($m=1$) & 5261 \AA \\
    Wedge prism apex angle & 25\degree.3 \\
    Physical dimension & 50 mm $\times$ 50 mm\\
    \hline
    \end{tabular}
    \end{table}

    \subsection{Imaging assembly}
    We selected a Canon RF 85 mm f/1.2 lens as the imaging lens. The lens has excellent imaging capability despite its small form factor and large FOV. The imaging lens is backed by Andor iKon-L 936 CCD, which serves as the CCD for SPECTR. 
    This Peltier-cooled CCD has a pixel array of $2048 \times 2048$ and a pixel size of 13.5 $\micron$ $\times$ 13.5 $\micron$. The designed FOV in imaging mode is 10\arcmin $\times$ 4\arcmin.6 with a plate scale of 0\arcsec.62/pixel. The gain and readout noise are 2.3~$\rm e^{-}$/ADU and 9~$\rm e^{-}$, respectively.

    \subsection{Slit monitoring and lamp systems of CIM}
\begin{figure*}\centering
    \includegraphics[width=0.90\textwidth]{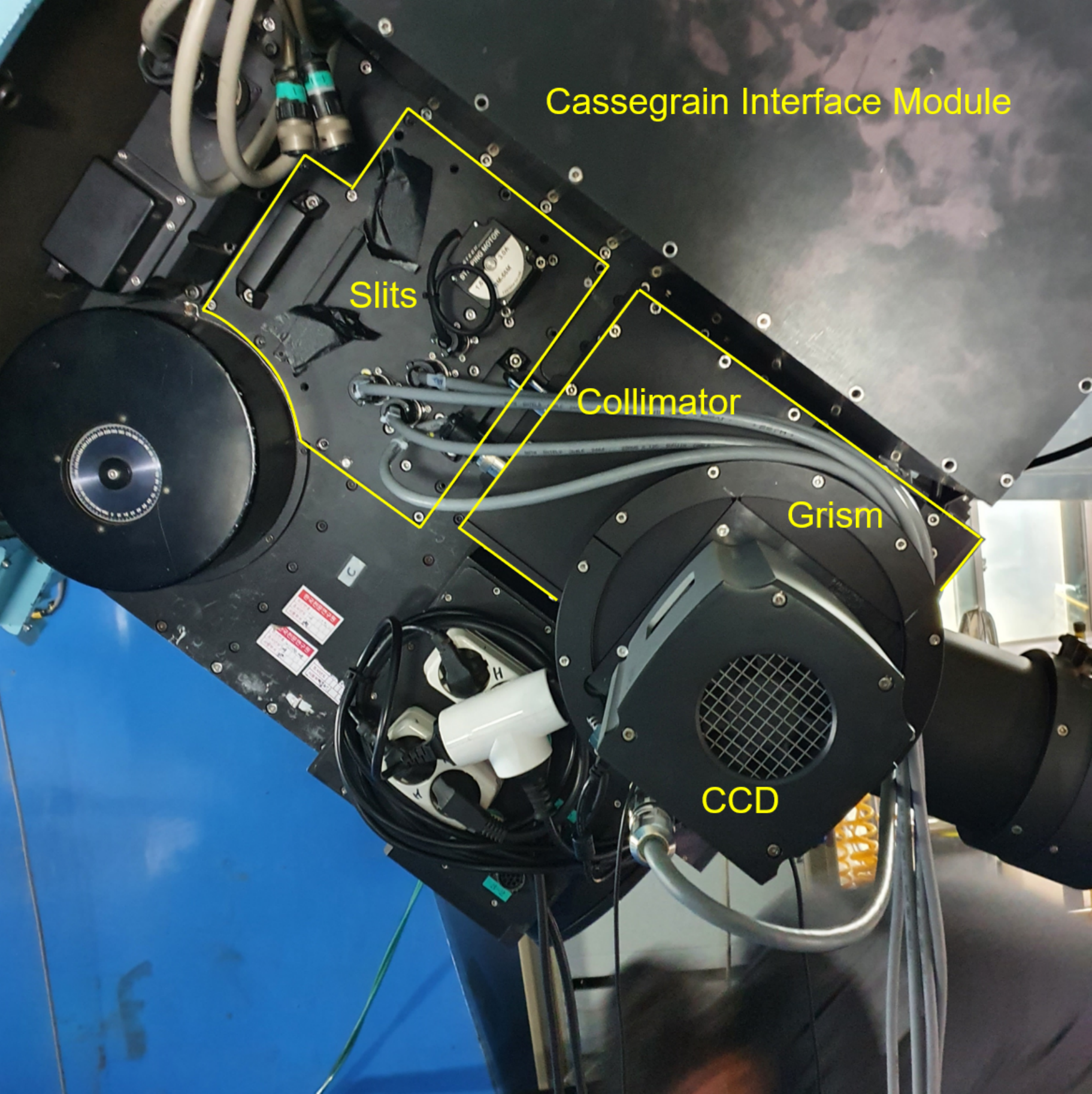}
    \caption{SPECTR installed on the CIM of BOAO 1.8 m telescope. The main parts of the instrument are highlighted in yellow.}
    \label{fig:SPECTR_photo}
    \end{figure*}

\begin{figure}[h!] \centering
    \includegraphics[width=0.47\textwidth]{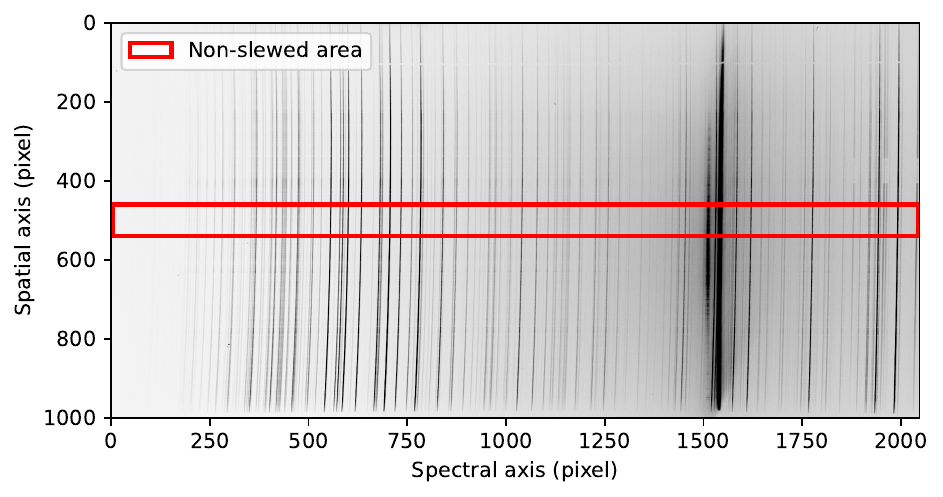}
    \caption{90\degree~rotated (clockwise) slewed FeAr lamp image. The red box indicates the 50\arcsec~of the spatial region that the FeAr lamp covers when it is centered in the light path without slewing.}
    \label{fig:fear}
    \end{figure}

    SPECTR installed on the CIM of the BOAO 1.8 m telescope is shown in Figure~\ref{fig:SPECTR_photo}.
    We designed SPECTR in a cartridge form factor, which can be installed and removed easily, and has compatibility with existing systems. Such a design has three advantages. First, it allows us to utilize the observing time before and after the transit event. 
    Transit observations typically require only a fraction of a night. Hence, a simple procedure to exchange SPECTR with BOES allows efficient usage of the allocated time.
    Second, SPECTR can utilize the slit-monitoring and auto-guiding systems within the CIM. The FOVs of each system are 3\arcmin~$\times$ 3\arcmin~and 40\arcmin~$\times$~40\arcmin, respectively. Precise guiding is crucial for transit observations lasting several hours to ensure that the whole point spread functions of both the target and the comparison star remain within the slit to avoid slit losses in the presence of changing seeing conditions.
    Although the slit-monitoring system has a limited FOV, it provides more precise guidance compared to the auto-guiding system. We utilize the auto-guiding system if no adequate star for guiding is available within this FOV.
    We checked the performance of each guiding system during test observations. The observations with the slit-monitoring system showed gradual positional drifts of approximately 1\arcsec.8 over three hours, while the auto-guiding system showed drifts of approximately 3\arcsec.6 shifts over four hours. 
    Third, it allows us to utilize the FeAr wavelength calibration lamp system in the CIM. Although SPECTR is designed to be integrated with the CIM, the existing calibration lamp system was not fully optimized for the SPECTR. When the calibration lamp is set to the optical axis, it only covers 50\arcsec~which is 8\% of the FOV of the SPECTR (red lines in Figure~\ref{fig:fear}). We modified the lamp system so that the lamp along the spatial axis covers the whole FOV of SPECTR. Figure~\ref{fig:fear} shows the resulting. The narrowest \#1 slit width of 1\arcsec.4~(R $\sim$ 1300) was selected to identify the wavelength solution and the rms value of the residuals after fitting was about 0.1 \AA.

\section{System Performance}
We tested the performance of SPECTR by observing transiting close-in exoplanets. The first test target was the well-known exoplanet HD~189733~b \citep{Bouchy2005}. 
The other was Qatar\mbox{-}8~b, discovered by the Qatar Exoplanet Survey \citep{Alsubai2013}. Our main purpose of the test was to validate the spectrophotometric stability and accuracy of SPECTR by comparing our results with existing spectroscopic light curves.

    \subsection{Observations}
    
    We observed with SPECTR one partial transit of HD~189733~b, starting at ingress, and one full transit of Qatar\mbox{-}8~b. A summary of the observations is provided in Table~\ref{t1}. Observations were carried out with the \#7 slit, which has a width of 15\arcsec, to minimize slit losses.
    Although the spectral resolution R corresponding to this slit width is approximately 150, the effective spectral resolution is determined by the seeing when the seeing is smaller than the slit width. The mean values of seeing for each observation were about 1.5\arcsec, resulting in a mean spectral resolution of approximately 1200.
    The auto-guiding system was used during the HD~189733~b observations, while the slit-monitoring system was used during the Qatar\mbox{-}8~b observations. All observations were carried out with the target and a comparison star simultaneously within the slit. We collected calibration data such as bias, dark, wavelength calibration, and sky flat in both observations.
    
    The partial transit of HD~189733~b was observed on October 14, 2021. The airmass changed during the observations from 1.03 to 1.61. We selected the comparison star HD~345459 from \citet{Bakos2006}. This comparison star had been chosen for photometry and spectrophotometry in \citet{Bakos2006} and \citet{Kasper2019}. The target and the comparison stars have similar spectral types and magnitudes, K2V with V = 7.64 and K0V with V = 8.09, respectively \citep{Nesterov1995, Hog2000, Gray2003, Koen2010}. A timeseries of 248 spectra of HD~189733 and the comparison star were obtained during in and out of transit with an exposure time of 15 seconds per frame. The cadence of the observations was 20 seconds, with five seconds of readout time, which was fixed throughout the entire test observation.

\begin{figure} \centering
    \includegraphics[width=0.47\textwidth]{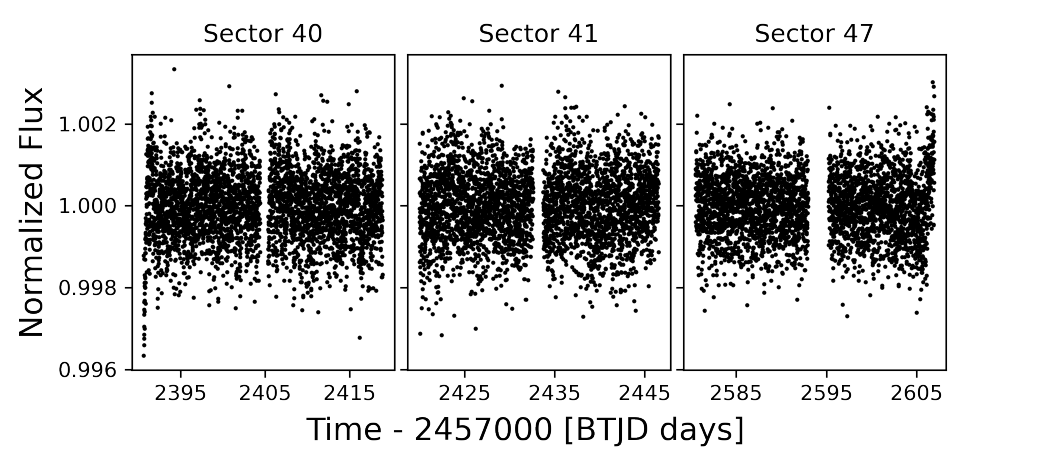}
    \caption{The \textit{TESS} light curves of the comparison star of Qatar\mbox{-}8, TYC 4387-01409-1 in \textit{TESS} Sectors 40, 41, and 47.}
    \label{fig:comp_TESS}
    \end{figure}

    One full transit of Qatar\mbox{-}8~b was observed on April 8, 2022. The airmass changed during the observations from 1.22 to 1.81. Qatar\mbox{-}8 is known to be a G0 V star \citep{Alsubai2019}. Qatar\mbox{-}8 is much fainter than HD~189733 and was chosen to test the spectrophotometric stability of SPECTR for a fainter system. We chose the comparison star TYC 4387-01409-1 (also known as TIC 103753216) from TIC v8 \citep{Stassun2019}. Although the spectral type of the comparison star is not known, we selected it based on its (B-V) color and checked its photometric variability from \textit{TESS} light curves. The \texttt{PDCSAP\_FLUX} (\citealt{Smith2012, Stumpe2014}) light curves of TYC 4387-01409-1 in \textit{TESS} Sectors 40, 41, and 47 appear stable at the rms = 0.8 ppt across all sectors, as shown in Figure~\ref{fig:comp_TESS}. 
    The target and the comparison star have similar magnitudes and colors, V = 11.71 with B-V = 0.534 and V = 11.65 with B-V = 0.573. A timeseries of 78 spectra of Qatar\mbox{-}8 and the comparison star were obtained in and out of transit with each exposure time of 300 seconds without readout time.

    \begin{table*}[]
    \centering
    \caption{Summary of observations.}
    \label{t1}
    {\scriptsize
    \begin{tabular}{ccccccccccc}
    \noalign{\smallskip}\noalign{\smallskip}\hline\hline
    \multirow{2}{*}{Target} & \multirow{2}{*}{Date of obs.} & \multicolumn{1}{c}{Obs. start} & \multicolumn{1}{c}{Obs. end} & \multicolumn{1}{c}{Transit start} & \multicolumn{1}{c}{Transit end} & \multirow{2}{*}{N$_\mathrm{exp}$} & \multirow{2}{*}{t$_\mathrm{exp}$ (s)} & \multirow{2}{*}{Airmass} & \multicolumn{1}{c}{Seeing} &\multirow{2}{*}{Moon illum.} \\
                            &                               & (UTC)                          & (UTC)                        & (UTC)                             & (UTC)                           &                       &                       &              &    (\arcsec)         &                              \\ \hline
    HD~189733~b & 2021-10-14 & 10:10 & 13:44 & 10:09 & 11:58 & 248 & 15 & 1.03 -- 1.61 & 1.17 -- 2.07 (1.43$^{a}$) & 0.64 \\
    Qatar\mbox{-}8~b & 2022-04-08 & 12:01 & 18:52 & 13:03 & 16:52 & 78 & 300 & 1.21 -- 1.81 & 1.33 -- 2.20 (1.57$^{a}$) & 0.43 \\
    \hline
    \end{tabular}
    \begin{tablenotes}
        \item{$^{a}$} Mean value of seeing 
    \end{tablenotes}
    }
    \end{table*}

    \subsection{Data reduction} \label{red}

    Data reduction was performed with standard IRAF \citep{IRAF1, IRAF2} procedures for bias, dark subtraction, illumination correction, aperture tracing, and wavelength calibration. From flat images of SPECTR, we found that uneven illumination does not significantly affect the results if the angular separation between the target and the comparison star is smaller than 7 arcminutes. For the HD~189733~b observations, the angular separation between the target and the comparison star was 8.72 arcminutes. Although we applied illumination correction to both targets, only the results for the HD~189733~b showed a significant difference.

    Aperture tracing and extraction were performed using the onedspec/apextract package. To confirm the aperture radii for both targets, we incrementally increased the radii from 3 to 15 pixels and extracted the aperture at each step. In this procedure, we fixed the background region as the pixel ranges from -30 to -20 and from 20 to 30 from the center to maintain consistency. Then we constructed white light curves of each aperture radius and compared the standard deviations of the out-of-transit light curves. The lowest scatters were found when we adopted aperture radii of 5.5 pixels for the HD~189733 and 11 pixels for the Qatar\mbox{-}8 observations. These aperture radii were fixed for the subsequent analysis. The difference in aperture radii between the two observations seems to arise from different seeing conditions.

\begin{figure} \centering
    \includegraphics[width=0.47\textwidth]{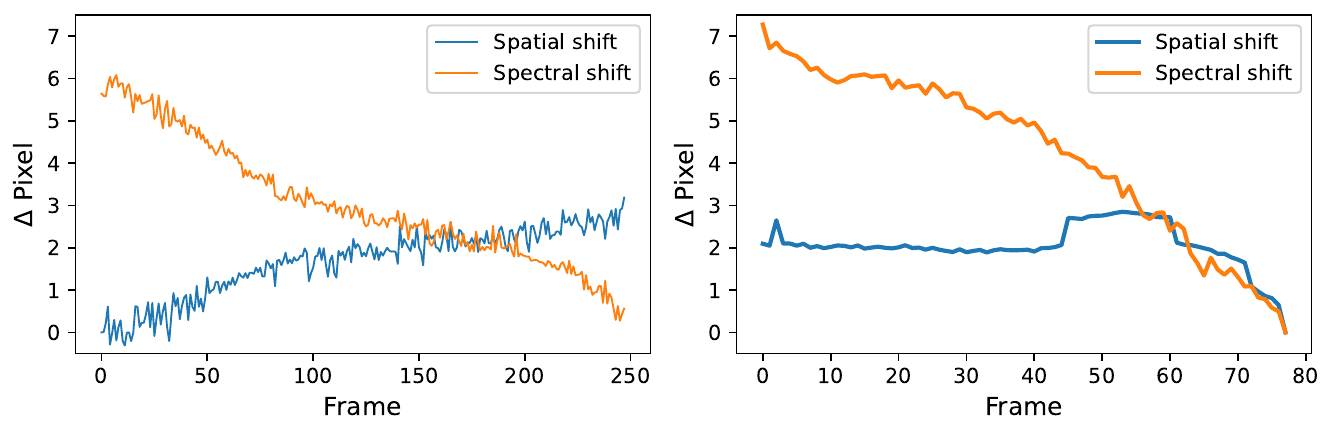}
    \caption{Pixel shifts over time in both spatial and spectral axes for HD 189733 b observations (left) and Qatar\mbox{-}8 b observations (right). The longer integration time for Qatar\mbox{-}8 b, guided with the slit-monitoring system, resulted in smoother shifts compared to HD 189733 b, which used only the auto-guiding system.}
    \label{fig:pixel_shift}
    \end{figure}

    Despite guiding, there were shifts along the spectral axis during the observations. We aligned the spectral axis of all spectra by comparing each spectrum with the reference spectrum using \texttt{scipy.optimize.minimize} function. Figure~\ref{fig:pixel_shift} shows the pixel shifts over time in both spatial and spectral axes. Given that the integration time for Qatar\mbox{-}8 b observations was approximately twice that of HD 189733 b observations, guiding with the slit-monitoring system appears significantly more effective. However, it is important to note that various confounding factors, such as atmospheric turbulence, the gravitational shifts depending on the position angle, and the position of the target in the sky, can impact guiding precision.
     
\begin{figure} \centering
    \includegraphics[width=0.47\textwidth]{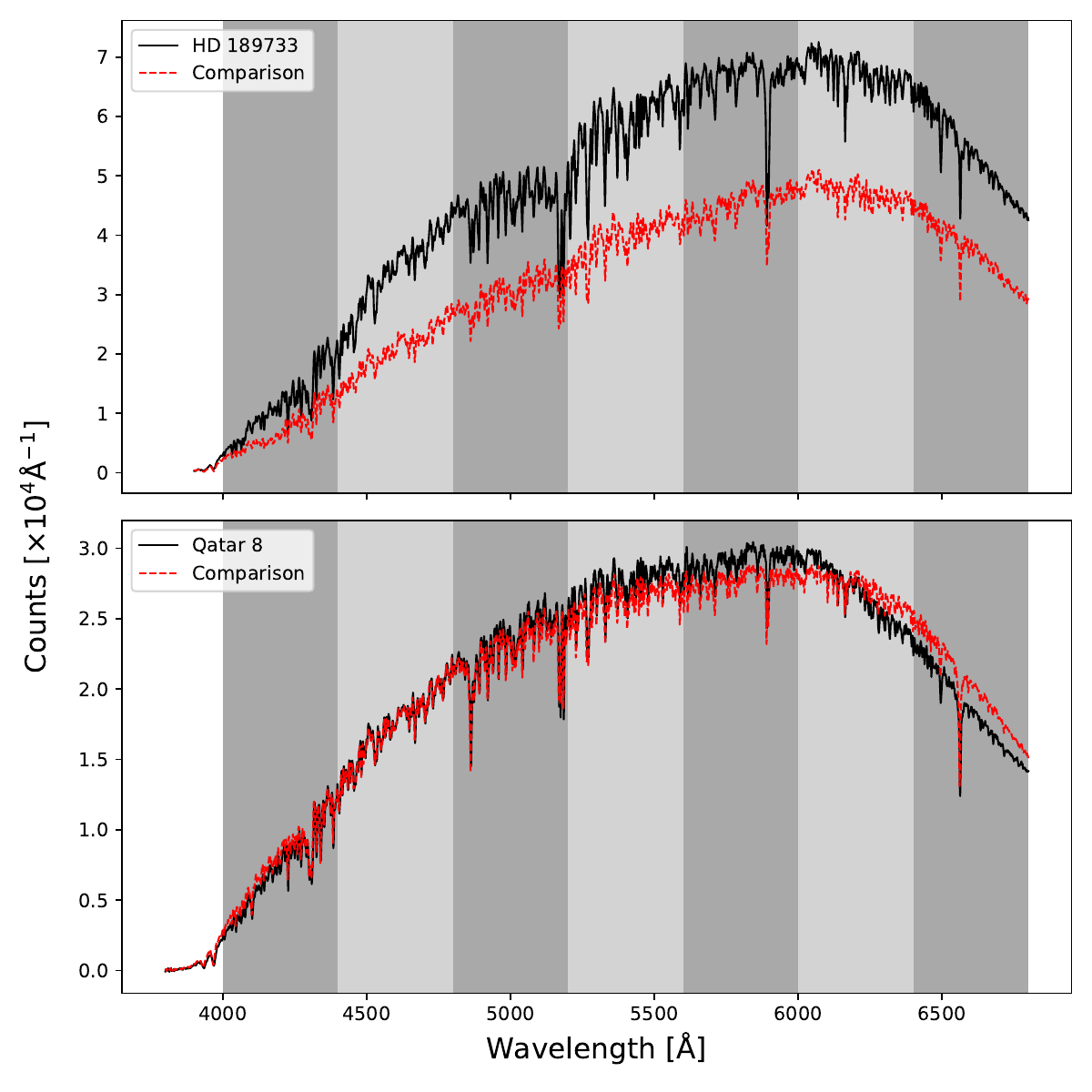}
    \caption{Extracted sample spectra of HD~189733 and Qatar\mbox{-}8 (black lines) along with their comparison stars (red lines). The wavelength bins used for constructing the spectroscopic light curves are indicated by grayed vertical regions.}
    \label{fig:spec}
    \end{figure}

    Figure~\ref{fig:spec} shows the extracted sample spectra of HD~189733 and Qatar\mbox{-}8, along with those of the comparison stars. Qatar\mbox{-}8 and its comparison star show quite similar spectral shapes, albeit with slightly different reported temperatures. The passbands of 400 \AA~width, used to construct seven spectroscopic light curves, are indicated by grayed vertical regions.

\begin{figure*} \centering
\includegraphics[width=0.47\textwidth]{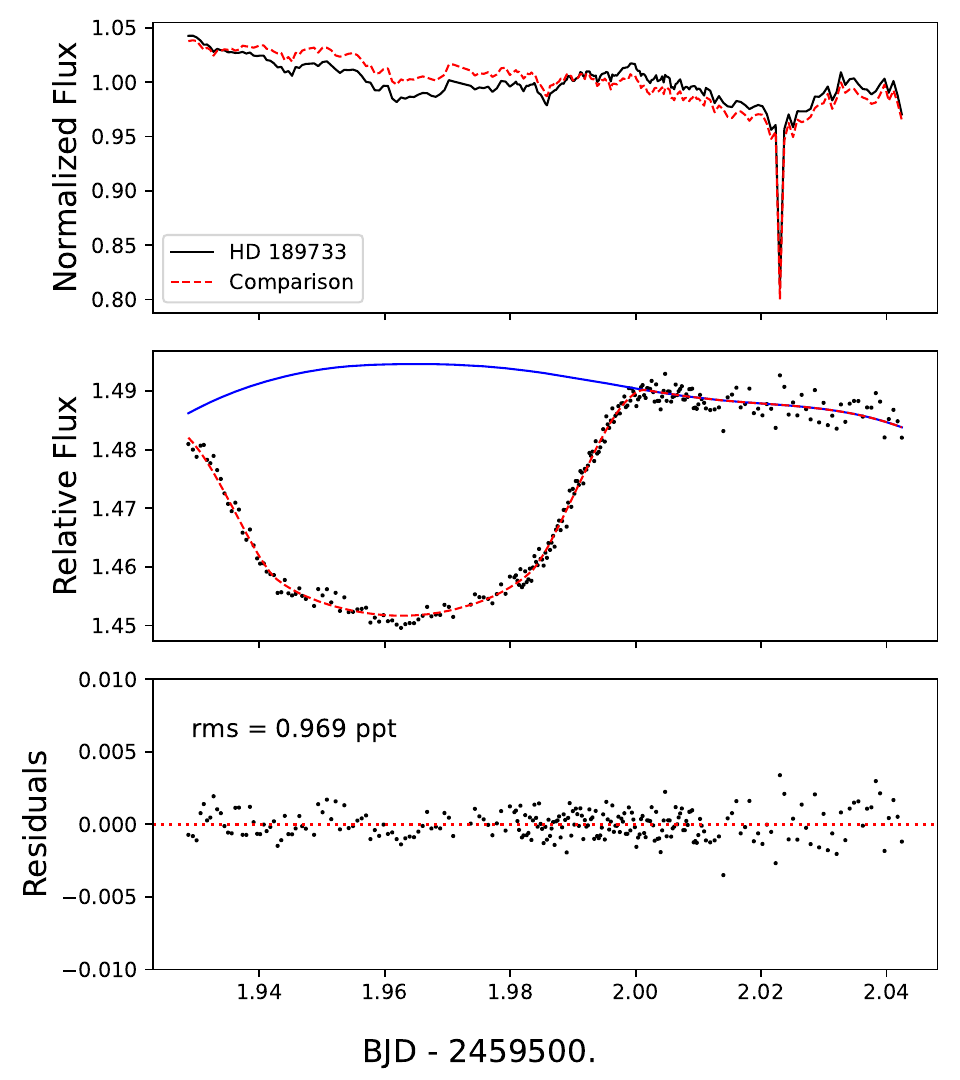}
\includegraphics[width=0.47\textwidth]{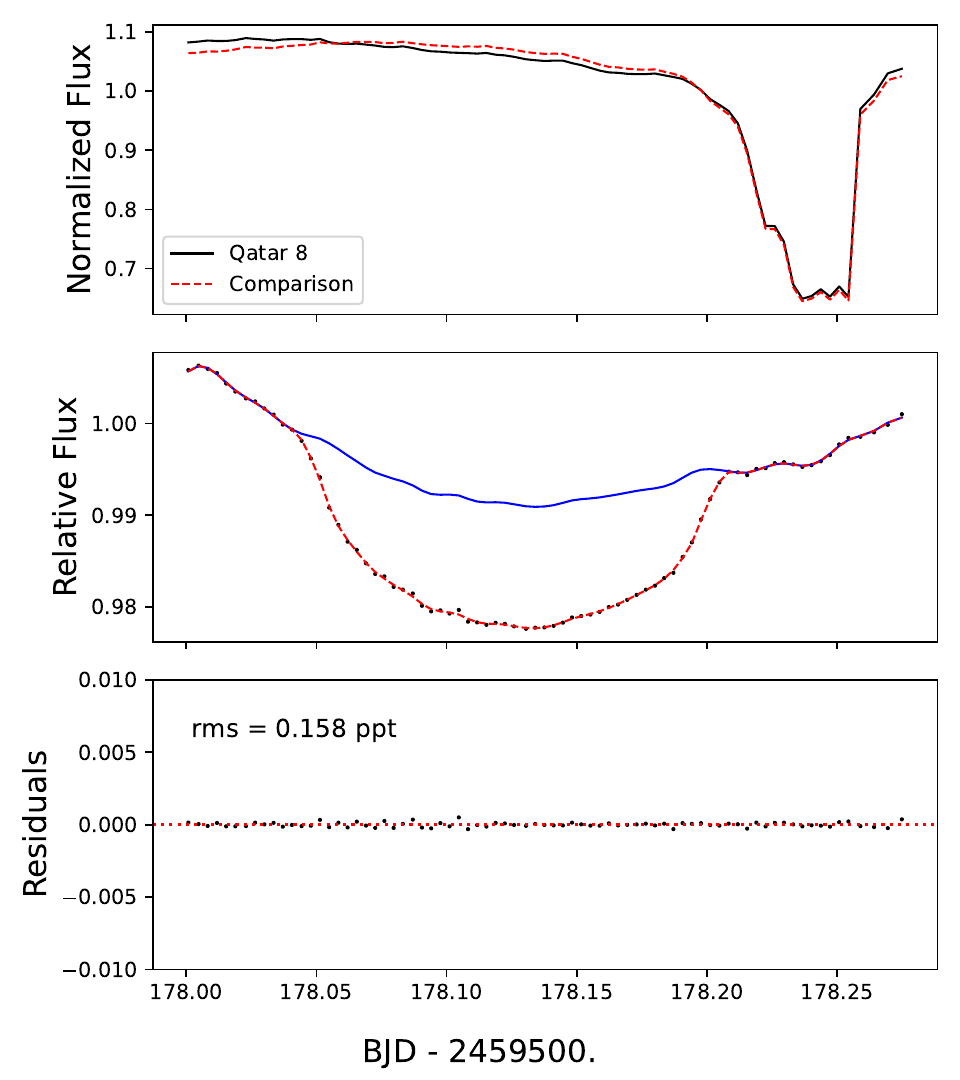}
\caption{The white light curves of HD~189733~b and Qatar\mbox{-}8~b. Upper panels: normalized raw light curves of targets and comparison stars. Middle panels: relative white light curves (black dots) and model fits (red line) for HD~189733~b and Qatar\mbox{-}8~b. The blue lines correspond to the detrended function produced by the GP analyses. Bottom panels: residuals after subtracting best-fitting.}
\label{fig:whitelc}
\end{figure*}

    \subsection{White and spectroscopic light curves}
    We integrated the flux across 4000 to 6800~\AA~to derive the white light curves for both targets, excluding wavelengths below 4000~\AA~ due to low SNRs.
    The white light curve of each target was obtained by dividing the integrated flux of the target by that of the comparison star. This removes the variations caused by the changing Earth's atmosphere as much as possible. Figure~\ref{fig:whitelc} shows the resulting white light curves, illustrating the observed flux variations during the transit events of HD~189733~b and Qatar\mbox{-}8~b, with model fits overlaying the data.
    
\begin{figure*}[!h] \centering
    \includegraphics[width=\columnwidth]{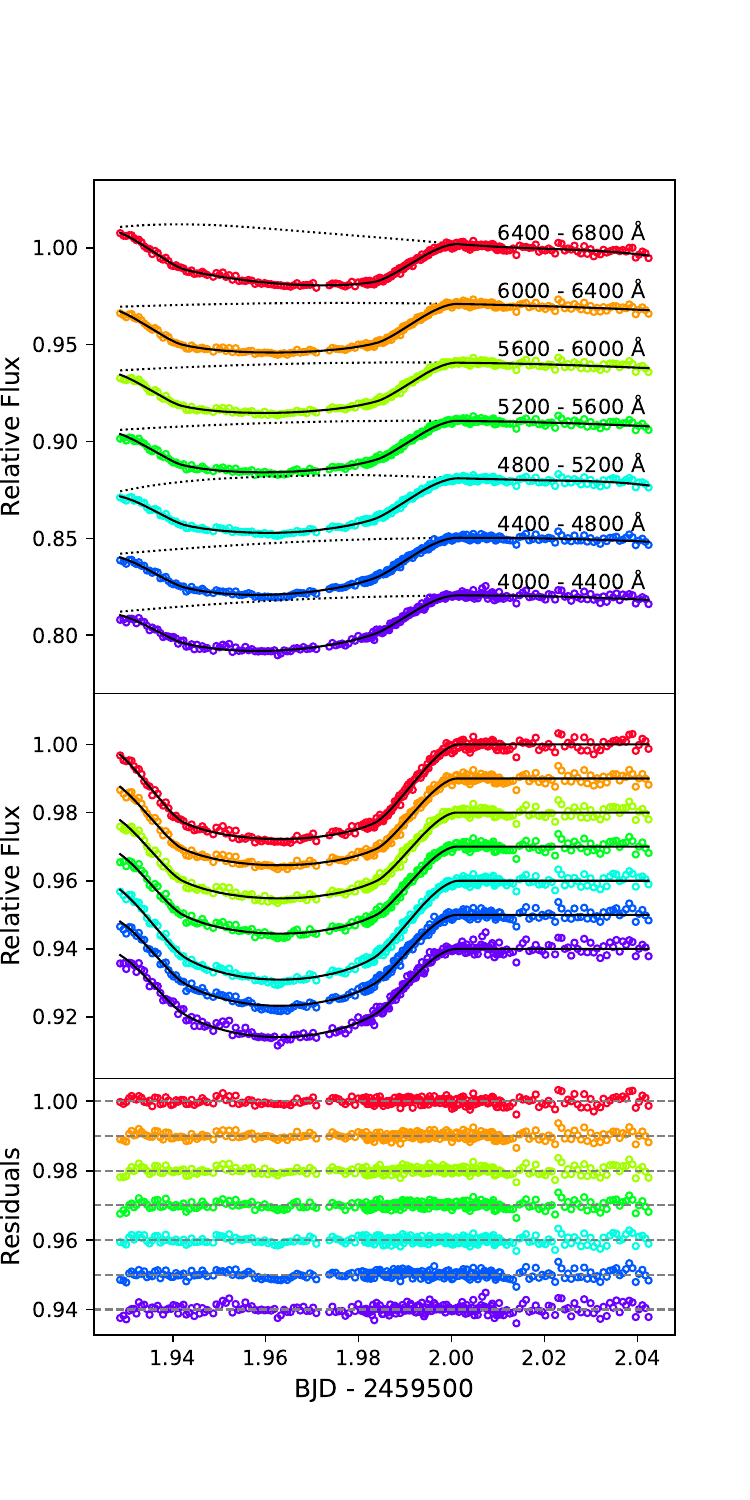}
    \includegraphics[width=\columnwidth]{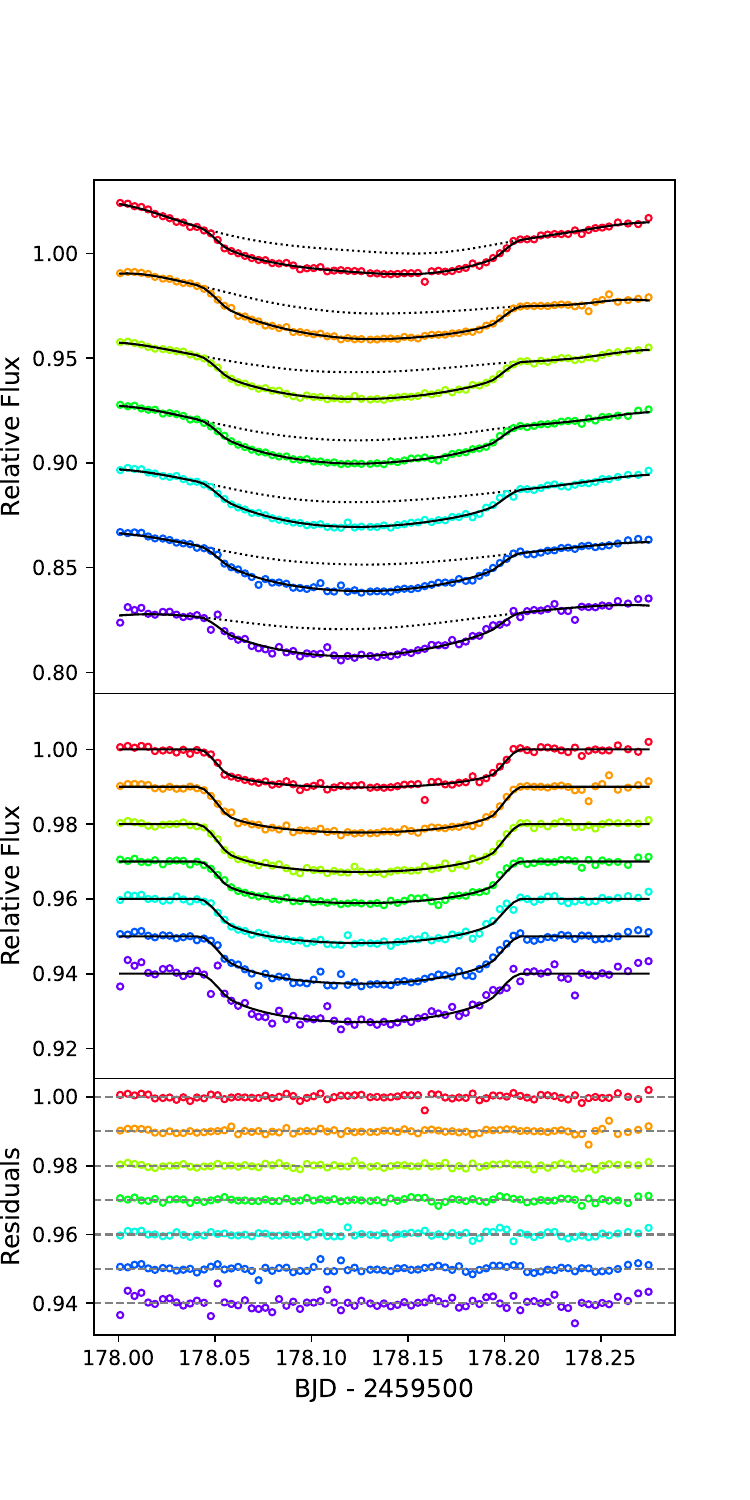}
    \caption{The spectroscopic light curves for HD~189733~b (left) and Qatar\mbox{-}8~b (right).
    The top panels display light curves of seven channels divided by the median value of out-of-transit phases.
    The dotted line of each channel shows the baseline fluxes modeled with GP.
    The middle and bottom panels show the detrended light curves with transit fits and their residuals, respectively.
    In all panels, the offsets were applied to the spectra for clarification.} 
    \label{fig:transmissionh}
    \end{figure*}

    We constructed the spectroscopic light curves by dividing the time-series spectra using uniform bins of 400~\AA~width.  
    Seven channels were extracted from 4,000~\AA~to 6,800~\AA~for each target and the light curve fits with the data from each channel are shown in Figure~\ref{fig:transmissionh}. 
    Spectroscopic light curves of both HD~189733~b and Qatar\mbox{-}8~b show similar shapes to the white light curves of each target before detrending. Since the systematics appear consistent across wavelength bins, we detrended the white light curves and the spectroscopic light curves in a similar manner.

    \subsection{Light curve fitting}

    We fitted the transit light curves using a quadratic limb-darkening transit model \citep{Mandel2002} implemented in \texttt{PyTransit} \citep{parviainen2015}. 
    Given that only a single transit event was observed for each target, we adopted the orbital parameters such as the period, semi-major axis, and inclination from previous studies \citep{Stassun2017, Alsubai2019}.
    Mid-transit time, planet-to-star radius ratio $R_{\rm p}/R_{\rm *}$, limb-darkening coefficients, and baseline trend were modeled using \texttt{PyTransit}.
    
    As shown in the middle panels of Figure~\ref{fig:whitelc}, the relative fluxes show both downward and upward parabolic trends. 
    To account for these trends over time, we applied the Gaussian Process (GP) with a Mat\'{e}rn 3/2 kernel, implemented in \texttt{celerite} \citep{Foreman2017}.
    The transit parameters and baseline trends were determined using a Bayesian framework.
    The baseline trends were derived by dividing the fluxes by the transit models and subsequently fitting them with the GP.
    
    We set a normal prior for all parameters, except for $R_{\rm p}/R_{\rm *}$, which had a uniform prior range between 0.01 and 0.2.
    The mid-transit time prior was set using a normal distribution with a standard deviation of 0.01 days. For HD 189733 b, the center value was 1.96368 days, and for Qatar\mbox{-}8 b, the center value was 178.125434 days, both referenced to BJD - 2,459,500 days, and the limb-darkening coefficients were calculated using the Python Limb Darkening Tool Kit \citep[\texttt{PyLDTk};][]{Parviainen2015b} based on literature values for stellar parameters (\Tef, \lgg, and \feh) listed in Table \ref{t2}. These values were used as means and standard deviations for the normal priors. 
    
    Optimal parameters were initially found using the differential evolution optimization method \citep{de} implemented in \texttt{PyTransit}. This was followed by MCMC sampling with \texttt{emcee} \citep{emcee} to estimate the final parameters while fixing the baseline trends.
    The mid-transit times for each object were fixed for the spectroscopic light curves based on the white light curve estimates.
    The limb-darkening coefficients remained consistent with the \texttt{PyLDTK} calculations after parameter estimation.

    The best-fit models for the white light curves of HD~189733~b and Qatar\mbox{-}8~b are shown in Figure~\ref{fig:whitelc}. 
    The summary of the best-fit and fixed parameters is listed in Table \ref{t2}. 
    The rms values of the residuals after the light curve model fitting are 0.969 ppt for HD~189733~b and 0.158 ppt for Qatar\mbox{-}8~b. 

\begin{figure*} \centering
    \includegraphics[width=0.47\textwidth]{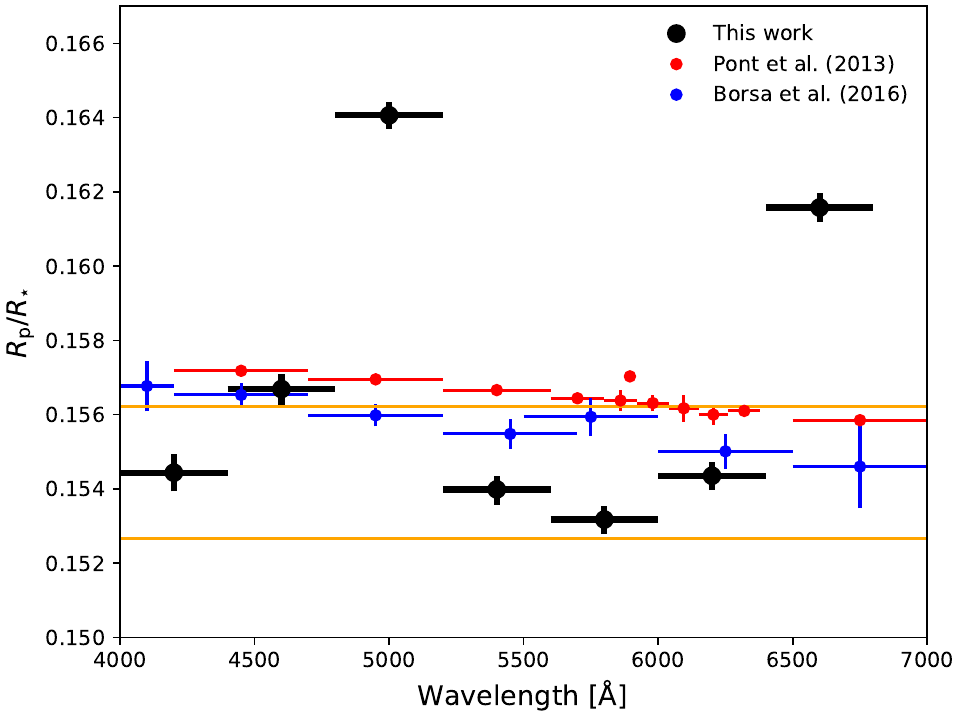}
    \includegraphics[width=0.47\textwidth]{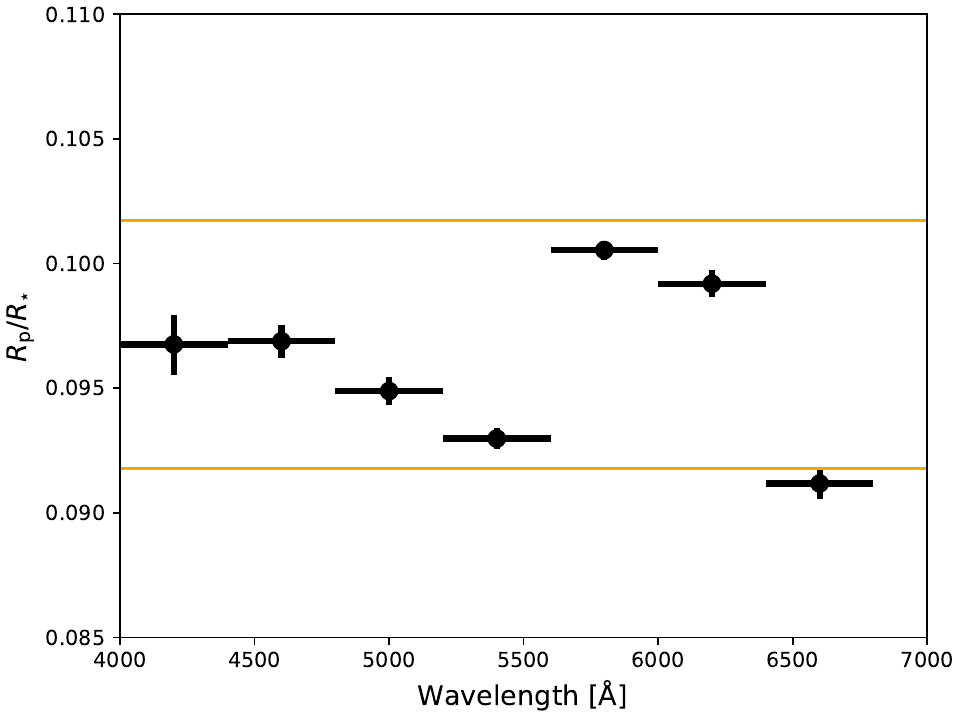}
    \caption{Transmission spectra of HD~189733~b (left) and Qatar\mbox{-}8~b (right) fitted using PyTransit. The overplotted points on the left panel are HST observations (red) by \citet{Pont2013} and HARPS observations (blue) by \citet{Borsa2016}.
    Two horizontal yellow lines of each panel indicate $\pm$5 atmospheric scale heights from a median of the transmission spectrum.}
    \label{fig:radii1}
    \end{figure*}

    We tabulate the spectral radius ratio from the spectroscopic light curve fits in Table \ref{t5}. Our transmission spectra along with those of \citet{Pont2013} and \citet{Borsa2016} are shown in Figure~\ref{fig:radii1}. The $R_{\rm p}/R_{\rm *}$ values for HD~189733~b determined in this work show a general agreement with those reported in previous studies, although two points deviate significantly from the general trend. This discrepancy may have been caused by the effects of the absence of data before ingress. 
    For Qatar\mbox{-}8~b, the $R_{\rm p}/R_{\rm *}$ values show a relatively smooth trend across the wavelengths with small increases in the 5800 \AA~and 6200 \AA~bins. 
    By additional monitoring of these targets, we aim to confirm whether these anomalies are indeed real. The mean rms values of the spectroscopic light curve fit residuals are 1.060 ppt for HD~189733~b and 0.843 ppt for Qatar\mbox{-}8~b.

    To illustrate how sensitive our transmission spectra are to potential atmospheric features in these two planets, we calculated the atmospheric scale heights of our targets using the equation from \citet{Etangs2008}, under the assumption of hydrogen-dominated atmospheres with a mean molecular weight ($\mu_{m}=2.37$). The atmospheric temperatures were taken to be equal to the equilibrium temperatures of the planets, and we used the known radii and masses of the planets. For HD~189733~b ($R_{\rm *} = 0.776~R_{\rm Sun}$ from \citet{Stassun2019}), the calculated scale height is 192 km, which corresponds to 0.0004 $R_{\rm p}/R_{\rm *}$. For Qatar\mbox{-}8~b ($R_{\rm *} = 1.315~R_{\rm Sun}$ from \citet{Alsubai2019}), the calculated scale height is 911 km, which corresponds to 0.0010 $R_{\rm p}/R_{\rm *}$. Figure~\ref{fig:radii1} shows the amplitudes that would correspond to $\pm 5$ scale heights with respect to the median of the transmission spectrum for each planet. This illustrates that SPECTR is sensitive to that range of atmospheric heights

    \begin{table}
    \caption{Derived best-fit parameters and adopted parameter values for the white light curve fits of each target and for their atmospheric scale height calculations. 
    }
    \label{t2}
    \begin{tabular}{lcc}
    
    \noalign{\smallskip}\noalign{\smallskip}
    \hline\hline
    \multicolumn{3}{l}{Stellar parameters} \\
    \hline    
    Parameter        & HD~189733                 & Qatar\mbox{-}8 \\
    \hline
    $T_{\rm{eff}}$ (K)          & 5023  $^{a}$  & 5738  $^{d}$ \\
    $\log~g$ ($\rm{cm/s^{2}}$)  & 4.582 $^{a}$  & 4.214 $^{d}$ \\
    $\left[\rm Fe/H \right]$    & 0.02  $^{a}$  & 0.025 $^{d}$ \\ 
    $R_{\rm *}$ ($R_{\rm Sun}$) & 0.776  $^{a}$ & 1.315 $^{d}$  \\
 
    \hline\hline
    \multicolumn{3}{l}{Planet parameters} \\
    \hline
    Parameter                       & HD~189733~b                          & Qatar\mbox{-}8~b      \\
    \hline
    $R_{\rm p}$ ($R_{\rm Jup}$) & 1.13 $^{b}$ & 1.285 $^{d}$ \\
    $M_{\rm p}$ ($M_{\rm Jup}$) & 1.13 $^{b}$ & 0.371 $^{d}$ \\
    $T_{\rm eq}$ (K)            & 1209 $^{c}$ & 1457  $^{d}$ \\
    \hline

    \hline\hline
    \multicolumn{3}{l}{Orbital transit parameters} \\
    \hline    
    Parameter         & HD~189733~b              & Qatar\mbox{-}8~b                \\
    \hline
    $T_{\rm 0}$ (days)    & 1.963007  $\pm$ 0.000077 & 178.125274 $\pm$ 0.00011   \\
    $R_{\rm p}/R_{\rm *}$ & 0.163487  $\pm$ 0.00037  & 0.10203    $\pm$ 0.00012   \\
    $u_1$                 & 0.7206    $\pm$  0.0016  & 0.6313     $\pm$ 0.0012    \\
    $u_2$                 & 0.0508    $\pm$  0.0036  & 0.1069     $\pm$ 0.0025    \\
    $a/R_{\rm *}$         & 8.84       $^{b}$   & 7.761   $^{d}$       \\
    $P$ (days)            & 2.21857567 $^{b}$   & 3.71495 $^{d}$       \\
    $i$ (degrees)          & 85.71      $^{b}$   & 89.29   $^{d}$       \\
    $e$               & 0.0           & 0.0            \\
    \hline\noalign{\smallskip}
    \end{tabular}
    * All parameters except $T_{\rm 0}$, $R_{\rm p}/R_{\rm *}$, $u_1$, and $u_2$ were fixed
    $T_{0}$ : BJD - 2,459,500               \\
    $^{a}$ adopted from \citet{Stassun2019} \\
    $^{b}$ adopted from \citet{Stassun2017} \\
    $^{c}$ adopted from \citet{Addison2019} \\
    $^{d}$ adopted from \citet{Alsubai2019} 
    \end{table}

    \begin{table}
    \centering
    \caption{Radius ratio per spectrophotometric bin for HD~189733~b and Qatar\mbox{-}8~b.}
    \label{t5}
    \begin{tabular}{ccc}
    \noalign{\smallskip}\noalign{\smallskip}\hline\hline
    Wavelength   & \multicolumn{2}{c}{$R_{\rm p}/R_{\rm *}$}                \\
      (\AA)      & HD~189733~b            & Qatar\mbox{-}8~b           \\
    \hline
    
    4000 -- 4400  & 0.1544 $\pm$ 0.0005  & 0.0968 $\pm$ 0.0012 \\
    4400 -- 4800  & 0.1567 $\pm$ 0.0004  & 0.0969 $\pm$ 0.0007 \\
    4800 -- 5200  & 0.1641 $\pm$ 0.0004  & 0.0949 $\pm$ 0.0006 \\ 
    5200 -- 5600  & 0.1540 $\pm$ 0.0004  & 0.0930 $\pm$ 0.0004 \\
    5600 -- 6000  & 0.1532 $\pm$ 0.0004  & 0.1005 $\pm$ 0.0004 \\
    6000 -- 6400  & 0.1544 $\pm$ 0.0004  & 0.0992 $\pm$ 0.0006 \\
    6400 -- 6800  & 0.1616 $\pm$ 0.0004  & 0.0912 $\pm$ 0.0006 \\
    \hline
    \end{tabular}
    \end{table}

\bigskip
\section{Summary}

We present the instrumental design and spectrophotometric performance of the SPECtrophotometer for TRansmission spectroscopy of exoplanets (SPECTR) to be used with the BOAO 1.8 m telescope.
With its 10\arcmin~long-slit capability and spectrophotometrical stability, SPECTR will provide observational opportunities for 1 -- 2 m class telescopes to contribute to the understanding of exoplanetary atmospheres, particularly those transiting bright stars. While many of these targets will be discovered by \textit{TESS}, others will come from various surveys and instruments. 
The test observations showed that SPECTR can obtain precise spectroscopic light curves with a mean rms precision of 0.951 ppt. Although the analysis of the HD~189733~b and Qatar\mbox{-}8~b can be further improved, our preliminary results highlight the potential of SPECTR to perform atmospheric characterization of transiting exoplanets.
As the importance of exoplanet atmospheric research is increasing along with the increasing number of bright exoplanet systems, we hope to make a meaningful contribution to the characterization of exoplanets at visible wavelengths with the SPECTR and the BOAO 1.8 m telescope.

\section*{Acknowledgments}

We would like to express our sincere gratitude to the referee for their extensive valuable and insightful comments, which have greatly improved the quality and clarity of this paper. This paper is based on observations collected with the BOAO 1.8 m telescope. This research has made use of the NASA Exoplanet Archive, which is operated by the California Institute of Technology, under contract with the National Aeronautics and Space Administration under the Exoplanet Exploration Program.
B.-C. Lee acknowledges partial support by the KASI (Korea Astronomy and Space Science Institute) grant 2024-1-832-03 and acknowledges support by the National Research Foundation of Korea (NRF) grant funded by the Korea government (MSIT) (No-2021R1A2C1009501).
M.-G. Park was supported by the Basic Science Research Program through the National Research Foundation of Korea (NRF) funded by the Ministry of Education (No-2019R1I1A3A02062242 and No-2018R1A6A1A06024970) and KASI under the R\&D program supervised by the Ministry of Science, ICT and Future Planning.
J.-R. Koo was supported by the National Research Foundation of Korea (NRF) grant funded by the Korean government (MSIT; grant No-2022R1C1C2004102)

\textit{Software}: \textsf{numpy} \citep{Numpy},  \textsf{matplotlib} \citep{Matplotlib}, \textsf{Astropy} \citep{Astropy13, Astropy18}, \textsf{SciPy} \citep{scipy}

\bibliography{sample631}

\begin{thebibliography}{}
\expandafter\ifx\csname natexlab\endcsname\relax\def\natexlab#1{#1}\fi
\providecommand{\url}[1]{\href{#1}{#1}}
\providecommand{\dodoi}[1]{doi:~\href{http://doi.org/#1}{\nolinkurl{#1}}}
\providecommand{\doeprint}[1]{\href{http://ascl.net/#1}{\nolinkurl{http://ascl.net/#1}}}
\providecommand{\doarXiv}[1]{\href{https://arxiv.org/abs/#1}{\nolinkurl{https://arxiv.org/abs/#1}}}

\bibitem[{{Addison} {et~al.}(2019){Addison}, {Wright}, {Wittenmyer}, {Horner}, {Mengel}, {Johns}, {Marti}, {Nicholson}, {Soutter}, {Bowler}, {Crossfield}, {Kane}, {Kielkopf}, {Plavchan}, {Tinney}, {Zhang}, {Clark}, {Clerte}, {Eastman}, {Swift}, {Bottom}, {Muirhead}, {McCrady}, {Herzig}, {Hogstrom}, {Wilson}, {Sliski}, {Johnson}, {Wright}, {Johnson}, {Blake}, {Riddle}, {Lin}, {Cornachione}, {Bedding}, {Stello}, {Huber}, {Marsden}, \& {Carter}}]{Addison2019}
{Addison}, B., {Wright}, D.~J., {Wittenmyer}, R.~A., {et~al.} 2019, \pasp, 131, 115003

\bibitem[{{Ahrer} {et~al.}(2022){Ahrer}, {Wheatley}, {Kirk}, {Gandhi}, {King}, \& {Louden}}]{Ahrer22}
{Ahrer}, E., {Wheatley}, P.~J., {Kirk}, J., {et~al.} 2022, \mnras, 510, 4857

\bibitem[{{Alderson} {et~al.}(2020){Alderson}, {Kirk}, {L{\'o}pez-Morales}, {Wheatley}, {Skillen}, {Henry}, {McGruder}, {Brogi}, {Louden}, \& {King}}]{Alderson20}
{Alderson}, L., {Kirk}, J., {L{\'o}pez-Morales}, M., {et~al.} 2020, \mnras, 497, 5182

\bibitem[{{Allen} {et~al.}(2022){Allen}, {Espinoza}, {Jord{\'a}n}, {L{\'o}pez-Morales}, {Apai}, {Rackham}, {Kirk}, {Osip}, {Weaver}, {McGruder}, {Ceballos}, {Reggiani}, {Brahm}, {Rodler}, {Lewis}, \& {Fraine}}]{Allen2022}
{Allen}, N.~H., {Espinoza}, N., {Jord{\'a}n}, A., {et~al.} 2022, \aj, 164, 153

\bibitem[{{Alsubai} {et~al.}(2019){Alsubai}, {Tsvetanov}, {Pyrzas}, {Latham}, {Bieryla}, {Eastman}, {Mislis}, {Esquerdo}, {Southworth}, {Mancini}, {Esamdin}, {Liu}, {Ma}, {Bretton}, {Pall{\'e}}, {Murgas}, {Vilchez}, {Parviainien}, {Monta{\~n}es-Rodriguez}, {Narita}, {Fukui}, {Kusakabe}, {Tamura}, {Barkaoui}, {Pozuelos}, {Gillon}, {Jehin}, {Benkhaldoun}, \& {Daassou}}]{Alsubai2019}
{Alsubai}, K., {Tsvetanov}, Z.~I., {Pyrzas}, S., {et~al.} 2019, \aj, 157, 224

\bibitem[{{Alsubai} {et~al.}(2013){Alsubai}, {Parley}, {Bramich}, {Horne}, {Collier Cameron}, {West}, {Sorensen}, {Pollacco}, {Smith}, \& {Fors}}]{Alsubai2013}
{Alsubai}, K.~A., {Parley}, N.~R., {Bramich}, D.~M., {et~al.} 2013, \actaa, 63, 465

\bibitem[{{Arcangeli} {et~al.}(2018){Arcangeli}, {D{\'e}sert}, {Line}, {Bean}, {Parmentier}, {Stevenson}, {Kreidberg}, {Fortney}, {Mansfield}, \& {Showman}}]{arc18}
{Arcangeli}, J., {D{\'e}sert}, J.-M., {Line}, M.~R., {et~al.} 2018, \apjl, 855, L30

\bibitem[{{Astropy Collaboration} {et~al.}(2013){Astropy Collaboration}, {Robitaille}, {Tollerud}, {Greenfield}, {Droettboom}, {Bray}, {Aldcroft}, {Davis}, {Ginsburg}, {Price-Whelan}, {Kerzendorf}, {Conley}, {Crighton}, {Barbary}, {Muna}, {Ferguson}, {Grollier}, {Parikh}, {Nair}, {Unther}, {Deil}, {Woillez}, {Conseil}, {Kramer}, {Turner}, {Singer}, {Fox}, {Weaver}, {Zabalza}, {Edwards}, {Azalee Bostroem}, {Burke}, {Casey}, {Crawford}, {Dencheva}, {Ely}, {Jenness}, {Labrie}, {Lim}, {Pierfederici}, {Pontzen}, {Ptak}, {Refsdal}, {Servillat}, \& {Streicher}}]{Astropy13}
{Astropy Collaboration}, {Robitaille}, T.~P., {Tollerud}, E.~J., {et~al.} 2013, \aap, 558, A33

\bibitem[{{Astropy Collaboration} {et~al.}(2018){Astropy Collaboration}, {Price-Whelan}, {Sip{\H{o}}cz}, {G{\"u}nther}, {Lim}, {Crawford}, {Conseil}, {Shupe}, {Craig}, {Dencheva}, {Ginsburg}, {VanderPlas}, {Bradley}, {P{\'e}rez-Su{\'a}rez}, {de Val-Borro}, {Aldcroft}, {Cruz}, {Robitaille}, {Tollerud}, {Ardelean}, {Babej}, {Bach}, {Bachetti}, {Bakanov}, {Bamford}, {Barentsen}, {Barmby}, {Baumbach}, {Berry}, {Biscani}, {Boquien}, {Bostroem}, {Bouma}, {Brammer}, {Bray}, {Breytenbach}, {Buddelmeijer}, {Burke}, {Calderone}, {Cano Rodr{\'\i}guez}, {Cara}, {Cardoso}, {Cheedella}, {Copin}, {Corrales}, {Crichton}, {D'Avella}, {Deil}, {Depagne}, {Dietrich}, {Donath}, {Droettboom}, {Earl}, {Erben}, {Fabbro}, {Ferreira}, {Finethy}, {Fox}, {Garrison}, {Gibbons}, {Goldstein}, {Gommers}, {Greco}, {Greenfield}, {Groener}, {Grollier}, {Hagen}, {Hirst}, {Homeier}, {Horton}, {Hosseinzadeh}, {Hu}, {Hunkeler}, {Ivezi{\'c}}, {Jain}, {Jenness}, {Kanarek}, {Kendrew}, {Kern}, {Kerzendorf}, {Khvalko}, {King}, {Kirkby}, {Kulkarni},
  {Kumar}, {Lee}, {Lenz}, {Littlefair}, {Ma}, {Macleod}, {Mastropietro}, {McCully}, {Montagnac}, {Morris}, {Mueller}, {Mumford}, {Muna}, {Murphy}, {Nelson}, {Nguyen}, {Ninan}, {N{\"o}the}, {Ogaz}, {Oh}, {Parejko}, {Parley}, {Pascual}, {Patil}, {Patil}, {Plunkett}, {Prochaska}, {Rastogi}, {Reddy Janga}, {Sabater}, {Sakurikar}, {Seifert}, {Sherbert}, {Sherwood-Taylor}, {Shih}, {Sick}, {Silbiger}, {Singanamalla}, {Singer}, {Sladen}, {Sooley}, {Sornarajah}, {Streicher}, {Teuben}, {Thomas}, {Tremblay}, {Turner}, {Terr{\'o}n}, {van Kerkwijk}, {de la Vega}, {Watkins}, {Weaver}, {Whitmore}, {Woillez}, {Zabalza}, \& {Astropy Contributors}}]{Astropy18}
{Astropy Collaboration}, {Price-Whelan}, A.~M., {Sip{\H{o}}cz}, B.~M., {et~al.} 2018, \aj, 156, 123

\bibitem[{{Bakos} {et~al.}(2006){Bakos}, {Knutson}, {Pont}, {Moutou}, {Charbonneau}, {Shporer}, {Bouchy}, {Everett}, {Hergenrother}, {Latham}, {Mayor}, {Mazeh}, {Noyes}, {Queloz}, {P{\'a}l}, \& {Udry}}]{Bakos2006}
{Bakos}, G.~{\'A}., {Knutson}, H., {Pont}, F., {et~al.} 2006, \apj, 650, 1160

\bibitem[{{Bixel} {et~al.}(2019){Bixel}, {Rackham}, {Apai}, {Espinoza}, {L{\'o}pez-Morales}, {Osip}, {Jord{\'a}n}, {McGruder}, \& {Weaver}}]{Bixel19}
{Bixel}, A., {Rackham}, B.~V., {Apai}, D., {et~al.} 2019, \aj, 157, 68

\bibitem[{{Borsa} {et~al.}(2016){Borsa}, {Rainer}, \& {Poretti}}]{Borsa2016}
{Borsa}, F., {Rainer}, M., \& {Poretti}, E. 2016, \aap, 590, A84

\bibitem[{{Borsa} {et~al.}(2021){Borsa}, {Allart}, {Casasayas-Barris}, {Tabernero}, {Zapatero Osorio}, {Cristiani}, {Pepe}, {Rebolo}, {Santos}, {Adibekyan}, {Bourrier}, {Demangeon}, {Ehrenreich}, {Pall{\'e}}, {Sousa}, {Lillo-Box}, {Lovis}, {Micela}, {Oshagh}, {Poretti}, {Sozzetti}, {Allende Prieto}, {Alibert}, {Amate}, {Benz}, {Bouchy}, {Cabral}, {Dekker}, {D'Odorico}, {Di Marcantonio}, {Figueira}, {Genova Santos}, {Gonz{\'a}lez Hern{\'a}ndez}, {Lo Curto}, {Manescau}, {Martins}, {M{\'e}gevand}, {Mehner}, {Molaro}, {Nunes}, {Riva}, {Su{\'a}rez Mascare{\~n}o}, {Udry}, \& {Zerbi}}]{Borsa_ESPRESSO}
{Borsa}, F., {Allart}, R., {Casasayas-Barris}, N., {et~al.} 2021, \aap, 645, A24

\bibitem[{{Bouchy} {et~al.}(2005){Bouchy}, {Udry}, {Mayor}, {Moutou}, {Pont}, {Iribarne}, {da Silva}, {Ilovaisky}, {Queloz}, {Santos}, {S{\'e}gransan}, \& {Zucker}}]{Bouchy2005}
{Bouchy}, F., {Udry}, S., {Mayor}, M., {et~al.} 2005, \aap, 444, L15

\bibitem[{{Charbonneau} {et~al.}(2002){Charbonneau}, {Brown}, {Noyes}, \& {Gilliland}}]{cha02}
{Charbonneau}, D., {Brown}, T.~M., {Noyes}, R.~W., \& {Gilliland}, R.~L. 2002, \apj, 568, 377

\bibitem[{{Crossfield} {et~al.}(2013){Crossfield}, {Barman}, {Hansen}, \& {Howard}}]{Crossfield_MOSFIRE}
{Crossfield}, I. J.~M., {Barman}, T., {Hansen}, B. M.~S., \& {Howard}, A.~W. 2013, \aap, 559, A33

\bibitem[{{Deming} {et~al.}(2013){Deming}, {Wilkins}, {McCullough}, {Burrows}, {Fortney}, {Agol}, {Dobbs-Dixon}, {Madhusudhan}, {Crouzet}, {Desert}, {Gilliland}, {Haynes}, {Knutson}, {Line}, {Magic}, {Mandell}, {Ranjan}, {Charbonneau}, {Clampin}, {Seager}, \& {Showman}}]{dem13}
{Deming}, D., {Wilkins}, A., {McCullough}, P., {et~al.} 2013, \apj, 774, 95

\bibitem[{{Espinoza} {et~al.}(2019){Espinoza}, {Rackham}, {Jord{\'a}n}, {Apai}, {L{\'o}pez-Morales}, {Osip}, {Grimm}, {Hoeijmakers}, {Wilson}, {Bixel}, {McGruder}, {Rodler}, {Weaver}, {Lewis}, {Fortney}, \& {Fraine}}]{Espinoza19}
{Espinoza}, N., {Rackham}, B.~V., {Jord{\'a}n}, A., {et~al.} 2019, \mnras, 482, 2065

\bibitem[{{Foreman-Mackey} {et~al.}(2017){Foreman-Mackey}, {Agol}, {Ambikasaran}, \& {Angus}}]{Foreman2017}
{Foreman-Mackey}, D., {Agol}, E., {Ambikasaran}, S., \& {Angus}, R. 2017, \aj, 154, 220

\bibitem[{{Foreman-Mackey} {et~al.}(2013){Foreman-Mackey}, {Hogg}, {Lang}, \& {Goodman}}]{emcee}
{Foreman-Mackey}, D., {Hogg}, D.~W., {Lang}, D., \& {Goodman}, J. 2013, \pasp, 125, 306

\bibitem[{{Gray} {et~al.}(2003){Gray}, {Corbally}, {Garrison}, {McFadden}, \& {Robinson}}]{Gray2003}
{Gray}, R.~O., {Corbally}, C.~J., {Garrison}, R.~F., {McFadden}, M.~T., \& {Robinson}, P.~E. 2003, \aj, 126, 2048

\bibitem[{{Han} {et~al.}(2010){Han}, {Lee}, {Kim}, {Mkrtichian}, {Hatzes}, \& {Valyavin}}]{Han2010}
{Han}, I., {Lee}, B.~C., {Kim}, K.~M., {et~al.} 2010, \aap, 509, A24

\bibitem[{{H{\o}g} {et~al.}(2000){H{\o}g}, {Fabricius}, {Makarov}, {Urban}, {Corbin}, {Wycoff}, {Bastian}, {Schwekendiek}, \& {Wicenec}}]{Hog2000}
{H{\o}g}, E., {Fabricius}, C., {Makarov}, V.~V., {et~al.} 2000, \aap, 355, L27

\bibitem[{{Huitson} {et~al.}(2017){Huitson}, {D{\'e}sert}, {Bean}, {Fortney}, {Stevenson}, \& {Bergmann}}]{Huitson_GMOS}
{Huitson}, C.~M., {D{\'e}sert}, J.~M., {Bean}, J.~L., {et~al.} 2017, \aj, 154, 95

\bibitem[{{Hunter}(2007)}]{Matplotlib}
{Hunter}, J.~D. 2007, Computing in Science and Engineering, 9, 90

\bibitem[{{JWST Transiting Exoplanet Community Early Release Science Team} {et~al.}(2023){JWST Transiting Exoplanet Community Early Release Science Team}, {Ahrer}, {Alderson}, {Batalha}, {Batalha}, {Bean}, {Beatty}, {Bell}, {Benneke}, {Berta-Thompson}, {Carter}, {Crossfield}, {Espinoza}, {Feinstein}, {Fortney}, {Gibson}, {Goyal}, {Kempton}, {Kirk}, {Kreidberg}, {L{\'o}pez-Morales}, {Line}, {Lothringer}, {Moran}, {Mukherjee}, {Ohno}, {Parmentier}, {Piaulet}, {Rustamkulov}, {Schlawin}, {Sing}, {Stevenson}, {Wakeford}, {Allen}, {Birkmann}, {Brande}, {Crouzet}, {Cubillos}, {Damiano}, {D{\'e}sert}, {Gao}, {Harrington}, {Hu}, {Kendrew}, {Knutson}, {Lagage}, {Leconte}, {Lendl}, {MacDonald}, {May}, {Miguel}, {Molaverdikhani}, {Moses}, {Murray}, {Nehring}, {Nikolov}, {Petit dit de la Roche}, {Radica}, {Roy}, {Stassun}, {Taylor}, {Waalkes}, {Wachiraphan}, {Welbanks}, {Wheatley}, {Aggarwal}, {Alam}, {Banerjee}, {Barstow}, {Blecic}, {Casewell}, {Changeat}, {Chubb}, {Col{\'o}n}, {Coulombe}, {Daylan}, {de Val-Borro},
  {Decin}, {Dos Santos}, {Flagg}, {France}, {Fu}, {Garc{\'\i}a Mu{\~n}oz}, {Gizis}, {Glidden}, {Grant}, {Heng}, {Henning}, {Hong}, {Inglis}, {Iro}, {Kataria}, {Komacek}, {Krick}, {Lee}, {Lewis}, {Lillo-Box}, {Lustig-Yaeger}, {Mancini}, {Mandell}, {Mansfield}, {Marley}, {Mikal-Evans}, {Morello}, {Nixon}, {Ortiz Ceballos}, {Piette}, {Powell}, {Rackham}, {Ramos-Rosado}, {Rauscher}, {Redfield}, {Rogers}, {Roman}, {Roudier}, {Scarsdale}, {Shkolnik}, {Southworth}, {Spake}, {Steinrueck}, {Tan}, {Teske}, {Tremblin}, {Tsai}, {Tucker}, {Turner}, {Valenti}, {Venot}, {Waldmann}, {Wallack}, {Zhang}, \& {Zieba}}]{JWST2023}
{JWST Transiting Exoplanet Community Early Release Science Team}, {Ahrer}, E.-M., {Alderson}, L., {et~al.} 2023, \nat, 614, 649

\bibitem[{{Kasper} {et~al.}(2019){Kasper}, {Cole}, {Gardner}, {Garver}, {Jarka}, {Kar}, {McGough}, {PeQueen}, {Rivera}, {Jang-Condell}, {Kobulnicky}, {Myers}, \& {Dale}}]{Kasper2019}
{Kasper}, D.~H., {Cole}, J.~L., {Gardner}, C.~N., {et~al.} 2019, \mnras, 483, 3781

\bibitem[{{Kim} {et~al.}(2002){Kim}, {Chun}, \& {Yoon}}]{Kim2002}
{Kim}, G.-M., {Chun}, M.-Y., \& {Yoon}, T.~S. 2002, Journal of Korean Astronomical Society, 35, 221

\bibitem[{{Kirk} {et~al.}(2020){Kirk}, {Alam}, {L{\'o}pez-Morales}, \& {Zeng}}]{Kirk_NIRSPEC}
{Kirk}, J., {Alam}, M.~K., {L{\'o}pez-Morales}, M., \& {Zeng}, L. 2020, \aj, 159, 115

\bibitem[{{Kirk} {et~al.}(2018){Kirk}, {Wheatley}, {Louden}, {Skillen}, {King}, {McCormac}, \& {Irwin}}]{Kirk18}
{Kirk}, J., {Wheatley}, P.~J., {Louden}, T., {et~al.} 2018, \mnras, 474, 876

\bibitem[{{Koen} {et~al.}(2010){Koen}, {Kilkenny}, {van Wyk}, \& {Marang}}]{Koen2010}
{Koen}, C., {Kilkenny}, D., {van Wyk}, F., \& {Marang}, F. 2010, \mnras, 403, 1949

\bibitem[{{Lecavelier Des Etangs} {et~al.}(2008){Lecavelier Des Etangs}, {Pont}, {Vidal-Madjar}, \& {Sing}}]{Etangs2008}
{Lecavelier Des Etangs}, A., {Pont}, F., {Vidal-Madjar}, A., \& {Sing}, D. 2008, \aap, 481, L83

\bibitem[{{Line} {et~al.}(2021){Line}, {Brogi}, {Bean}, {Gandhi}, {Zalesky}, {Parmentier}, {Smith}, {Mace}, {Mansfield}, {Kempton}, {Fortney}, {Shkolnik}, {Patience}, {Rauscher}, {D{\'e}sert}, \& {Wardenier}}]{Line_IGRINS}
{Line}, M.~R., {Brogi}, M., {Bean}, J.~L., {et~al.} 2021, \nat, 598, 580

\bibitem[{{Mandel} \& {Agol}(2002)}]{Mandel2002}
{Mandel}, K., \& {Agol}, E. 2002, \apjl, 580, L171

\bibitem[{{McGruder} {et~al.}(2020){McGruder}, {L{\'o}pez-Morales}, {Espinoza}, {Rackham}, {Apai}, {Jord{\'a}n}, {Osip}, {Alam}, {Bixel}, {Fortney}, {Henry}, {Kirk}, {Lewis}, {Rodler}, \& {Weaver}}]{McGruder2020}
{McGruder}, C.~D., {L{\'o}pez-Morales}, M., {Espinoza}, N., {et~al.} 2020, \aj, 160, 230

\bibitem[{{Nesterov} {et~al.}(1995){Nesterov}, {Kuzmin}, {Ashimbaeva}, {Volchkov}, {R{\"o}ser}, \& {Bastian}}]{Nesterov1995}
{Nesterov}, V.~V., {Kuzmin}, A.~V., {Ashimbaeva}, N.~T., {et~al.} 1995, \aaps, 110, 367

\bibitem[{{Nikolov} {et~al.}(2016){Nikolov}, {Sing}, {Gibson}, {Fortney}, {Evans}, {Barstow}, {Kataria}, \& {Wilson}}]{Nikolov_FORS2}
{Nikolov}, N., {Sing}, D.~K., {Gibson}, N.~P., {et~al.} 2016, \apj, 832, 191

\bibitem[{{Parviainen}(2015)}]{parviainen2015}
{Parviainen}, H. 2015, \mnras, 450, 3233

\bibitem[{{Parviainen} \& {Aigrain}(2015)}]{Parviainen2015b}
{Parviainen}, H., \& {Aigrain}, S. 2015, \mnras, 453, 3821

\bibitem[{{Pont} {et~al.}(2013){Pont}, {Sing}, {Gibson}, {Aigrain}, {Henry}, \& {Husnoo}}]{Pont2013}
{Pont}, F., {Sing}, D.~K., {Gibson}, N.~P., {et~al.} 2013, \mnras, 432, 2917

\bibitem[{{Rackham} {et~al.}(2017){Rackham}, {Espinoza}, {Apai}, {L{\'o}pez-Morales}, {Jord{\'a}n}, {Osip}, {Lewis}, {Rodler}, {Fraine}, {Morley}, \& {Fortney}}]{Rackham17}
{Rackham}, B., {Espinoza}, N., {Apai}, D., {et~al.} 2017, \apj, 834, 151

\bibitem[{{Sing} {et~al.}(2016){Sing}, {Fortney}, {Nikolov}, {Wakeford}, {Kataria}, {Evans}, {Aigrain}, {Ballester}, {Burrows}, {Deming}, {D{\'e}sert}, {Gibson}, {Henry}, {Huitson}, {Knutson}, {Lecavelier Des Etangs}, {Pont}, {Showman}, {Vidal-Madjar}, {Williamson}, \& {Wilson}}]{sin16}
{Sing}, D.~K., {Fortney}, J.~J., {Nikolov}, N., {et~al.} 2016, \nat, 529, 59

\bibitem[{{Smith} {et~al.}(2012){Smith}, {Stumpe}, {Van Cleve}, {Jenkins}, {Barclay}, {Fanelli}, {Girouard}, {Kolodziejczak}, {McCauliff}, {Morris}, \& {Twicken}}]{Smith2012}
{Smith}, J.~C., {Stumpe}, M.~C., {Van Cleve}, J.~E., {et~al.} 2012, \pasp, 124, 1000

\bibitem[{{Stassun} {et~al.}(2017){Stassun}, {Collins}, \& {Gaudi}}]{Stassun2017}
{Stassun}, K.~G., {Collins}, K.~A., \& {Gaudi}, B.~S. 2017, \aj, 153, 136

\bibitem[{{Stassun} {et~al.}(2018){Stassun}, {Oelkers}, {Pepper}, {Paegert}, {De Lee}, {Torres}, {Latham}, {Charpinet}, {Dressing}, {Huber}, {Kane}, {L{\'e}pine}, {Mann}, {Muirhead}, {Rojas-Ayala}, {Silvotti}, {Fleming}, {Levine}, \& {Plavchan}}]{Stassun2018}
{Stassun}, K.~G., {Oelkers}, R.~J., {Pepper}, J., {et~al.} 2018, \aj, 156, 102

\bibitem[{{Stassun} {et~al.}(2019){Stassun}, {Oelkers}, {Paegert}, {Torres}, {Pepper}, {De Lee}, {Collins}, {Latham}, {Muirhead}, {Chittidi}, {Rojas-Ayala}, {Fleming}, {Rose}, {Tenenbaum}, {Ting}, {Kane}, {Barclay}, {Bean}, {Brassuer}, {Charbonneau}, {Ge}, {Lissauer}, {Mann}, {McLean}, {Mullally}, {Narita}, {Plavchan}, {Ricker}, {Sasselov}, {Seager}, {Sharma}, {Shiao}, {Sozzetti}, {Stello}, {Vanderspek}, {Wallace}, \& {Winn}}]{Stassun2019}
{Stassun}, K.~G., {Oelkers}, R.~J., {Paegert}, M., {et~al.} 2019, \aj, 158, 138

\bibitem[{Storn \& Price(1997)}]{de}
Storn, R., \& Price, K.~V. 1997, J. Glob. Optim., 11, 341

\bibitem[{{Stumpe} {et~al.}(2014){Stumpe}, {Smith}, {Catanzarite}, {Van Cleve}, {Jenkins}, {Twicken}, \& {Girouard}}]{Stumpe2014}
{Stumpe}, M.~C., {Smith}, J.~C., {Catanzarite}, J.~H., {et~al.} 2014, \pasp, 126, 100

\bibitem[{{Tody}(1986)}]{IRAF1}
{Tody}, D. 1986, in Society of Photo-Optical Instrumentation Engineers (SPIE) Conference Series, Vol. 627, Instrumentation in astronomy VI, ed. D.~L. {Crawford}, 733

\bibitem[{{Tody}(1993)}]{IRAF2}
{Tody}, D. 1993, in Astronomical Society of the Pacific Conference Series, Vol.~52, Astronomical Data Analysis Software and Systems II, ed. R.~J. {Hanisch}, R.~J.~V. {Brissenden}, \& J.~{Barnes}, 173

\bibitem[{{van der Walt} {et~al.}(2011){van der Walt}, {Colbert}, \& {Varoquaux}}]{Numpy}
{van der Walt}, S., {Colbert}, S.~C., \& {Varoquaux}, G. 2011, Computing in Science and Engineering, 13, 22

\bibitem[{{Vidal-Madjar} {et~al.}(2003){Vidal-Madjar}, {Lecavelier des Etangs}, {D{\'e}sert}, {Ballester}, {Ferlet}, {H{\'e}brard}, \& {Mayor}}]{vid03}
{Vidal-Madjar}, A., {Lecavelier des Etangs}, A., {D{\'e}sert}, J.~M., {et~al.} 2003, \nat, 422, 143

\bibitem[{{Virtanen} {et~al.}(2020){Virtanen}, {Gommers}, {Oliphant}, {Haberland}, {Reddy}, {Cournapeau}, {Burovski}, {Peterson}, {Weckesser}, {Bright}, {van der Walt}, {Brett}, {Wilson}, {Millman}, {Mayorov}, {Nelson}, {Jones}, {Kern}, {Larson}, {Carey}, {Polat}, {Feng}, {Moore}, {VanderPlas}, {Laxalde}, {Perktold}, {Cimrman}, {Henriksen}, {Quintero}, {Harris}, {Archibald}, {Ribeiro}, {Pedregosa}, {van Mulbregt}, \& {SciPy 1. 0 Contributors}}]{scipy}
{Virtanen}, P., {Gommers}, R., {Oliphant}, T.~E., {et~al.} 2020, Nature Methods, 17, 261, \dodoi{10.1038/s41592-019-0686-2}

\end{thebibliography}
\bibliographystyle{aasjournal}

\end{document}